\PassOptionsToPackage{square,comma,numbers,sort&compress}{natbib}
\documentclass[12pt,singlespacing]{elsarticle}
\usepackage{graphicx}
\usepackage{amsmath}
\usepackage[table]{xcolor}
\usepackage{multirow}
\usepackage{amssymb}
\usepackage{setspace}
\usepackage{times}
\usepackage{natbib}
\usepackage{soul}
\usepackage{fullpage}
\usepackage{datetime}
\usepackage{caption}
\usepackage{subcaption}
\usepackage{placeins}
\usepackage{bm}
\usepackage{bm,upgreek}
\usepackage{amsbsy}
\usepackage[linesnumbered,ruled]{algorithm2e}
\usepackage{algpseudocode}
\usepackage[version=4]{mhchem}
\usepackage{amssymb}
\usepackage{mathtools}
\usepackage[T1]{fontenc}
\usepackage[title]{appendix}
\usepackage{booktabs}
\usepackage{adjustbox}
\usepackage{cleveref}
\usepackage{textcomp}
\usepackage[utf8]{inputenc}
\usepackage{amssymb}
\usepackage{multicol}
\usepackage{nomencl}
\makenomenclature
\usepackage{etoolbox}
\usepackage{wasysym}
\renewcommand{\nompreamble}{\begin{multicols}{2}}
\renewcommand{\nompostamble}{\end{multicols}}

\newcommand{\eref}[1]{(\ref{#1})}
\newcommand{\fref}[1]{Figure \ref{#1}}
\newcommand{\sref}[1]{Section \ref{#1}}
\newcommand{\rom}[1]{%
  \textup{\uppercase\expandafter{\romannumeral#1}}%
}

\SetKwInOut{Parameter}{parameter}

\begin{document}

\begin{frontmatter}
\title{\large Numerical Investigation of Critical Electrochemical Factors for Localized Corrosion using a Multi-species Reactive Transport Model}
\author[vu]{Xiangming Sun}
\author[osu]{Jayendran Srinivasan}
\author[uva]{Robert G. Kelly}
\author[vu]{Ravindra Duddu\corref{cor}}
\ead{ravindra.duddu@vanderbilt.edu}
\cortext[cor]{Corresponding author}
\address[vu]{Department of Civil and Environmental Engineering, Vanderbilt University, Nashville, TN 37235, USA}
\address[osu]{Fontana Corrosion Center, Department of Materials Science and Engineering, The Ohio State University, Columbus, OH 43210, USA}
\address[uva]{Center for Electrochemical Science and Engineering, Department of Materials Science and Engineering, University of Virginia, Charlottesville, VA 22904, USA}
\begin{abstract}
A multi-species reactive transport model based on the sequential non-iterative approach is employed to investigate two major stages involved in artificial pit experiments of stainless steel: I. Stable pitting under a salt film and II. Film-free dissolution that transitions to repassivation. Data of current density and electrical potential obtained from rapid polarization scans of pits with different depths are utilized to calibrate the elapsed time and the electrode kinetics of each stage. The local chemistry near the base of pits at different temperature and bulk concentrations is simulated to determine several critical electrochemical factors at saturation and repassivation.
\end{abstract}
\begin{keyword}
A. Stainless steel \sep B. Modeling studies \sep C. Pitting corrosion \sep C. Repassivation
\end{keyword}
\end{frontmatter}

\mbox{}

\nomenclature[01]{$\alpha_r$}{specific resistance ($\Omega\cdot$cm)}
\nomenclature[02]{$\zeta_{mn}$}{coefficient in Eq. \eref{eq:viscosityco}}
\nomenclature[03]{$\eta$}{viscosity coefficient}
\nomenclature[04]{$\kappa$}{electrode kinetics coefficient}
\nomenclature[05]{$\nu$}{stoichiometric coefficient}
\nomenclature[06]{$\tau$}{relative temperature (K)}
\nomenclature[07]{$\psi$}{electrical potential (V)}
\nomenclature[08]{$\omega$}{indicator parameter in Eq. \eref{eq:correctedconcentration}}
\nomenclature[09]{$a_i$}{coefficient in Eq. \eref{eq:viscosity}}
\nomenclature[10]{$C$}{corrected concentration (mol/m$^3$)}
\nomenclature[11]{$\widetilde{C}$}{uncorrected concentration (mol/m$^3$)}
\nomenclature[12]{$\hat{C}$}{concentration correction (mol/m$^3$)}
\nomenclature[13]{$C^{\infty}$}{far-field concentration (mol/m$^3$)}
\nomenclature[14]{$C_\text{Me}$}{concentration of metal cations (mol/m$^3$)}
\nomenclature[15]{$C_\text{sat}$}{saturation concentration (mol/m$^3$)}
\nomenclature[16]{$d$}{pit depth (m)}
\nomenclature[17]{$D$}{diffusion coefficient (m$^2$/s)}
\nomenclature[18]{$E_\text{app}$}{applied electrical potential (V)}
\nomenclature[19]{$E_\text{rp}$}{repassivation potential (V)}
\nomenclature[20]{$E_\text{T}$}{transition potential (V)}
\nomenclature[21]{$\dot{E}_\text{app}$}{polarization scan rate (V/s)}
\nomenclature[22]{$\Delta E_\text{IR}$}{IR drop (V)}
\nomenclature[23]{$F$}{Faraday's constant (C/mol)}
\nomenclature[24]{$f_\text{t/b}$}{flux of \ce{OH^-} (mol/m$^2$/s)}
\nomenclature[25]{$i_\text{a}$}{anodic current density (A/m$^2$)}
\nomenclature[26]{$i_\text{c,local}$}{local cathodic current density (A/m$^2$)}
\nomenclature[27]{$i_\text{L}$}{diffusion-limited current density (A/m$^2$)}
\nomenclature[28]{$i_\text{net}$}{net current density (A/m$^2$)}
\nomenclature[29]{$i_\text{rp}$}{repassivation current density (A/m$^2$)}
\nomenclature[30]{$\boldsymbol{J}$}{flux density (mol/m$^2$/s)}
\nomenclature[31]{$J_\text{Me}$}{flux of metal cations (mol/m$^2$/s)}
\nomenclature[32]{$K_r$}{reaction equilibrium constant}
\nomenclature[33]{$L$}{length of bulk domain (m)}
\nomenclature[34]{$n$}{number of species}
\nomenclature[35]{$\boldsymbol{n}$}{unit normal vector}
\nomenclature[36]{$n_r$}{number of reactions}
\nomenclature[37]{$p$}{pressure (Pa)}
\nomenclature[38]{$R$}{chemical reaction rate (mol/m$^3$/s)}
\nomenclature[39]{$R^*$}{universal gas constant (J/mol/K)}
\nomenclature[40]{$R_\text{b}$}{bulk resistance ($\Omega$)}
\nomenclature[41]{$R_l$}{electrolyte resistance ($\Omega$)}
\nomenclature[42]{$R_\text{p}$}{pit resistance ($\Omega$)}
\nomenclature[43]{$t$}{time (s)}
\nomenclature[44]{$T$}{temperature (K)}
\nomenclature[45]{$t_\text{II}$}{elapsed time of Stage II (s)}
\nomenclature[46]{$t_\text{d}$}{characteristic diffusion time (s)}
\nomenclature[47]{$\Delta t$}{time increment size (s)}
\nomenclature[48]{$W$}{width of bulk domain (m)}
\nomenclature[49]{$X_\text{ps}$}{pit stability product}
\nomenclature[50]{$z$}{charge number}
\nomenclature[51]{$\diameter$}{pit diameter (m)}

\printnomenclature[0.35in]

\section{Introduction}\label{sec:introduction}
Austenitic stainless steels are widely used in military mobile and fixed equipment and structures due to their outstanding resistance to uniform corrosion \citep{fossati2006corrosion,mccafferty2010introduction}. However, the main use of these equipment and structures is in harsh marine environments, which can be acidic and usually contain high concentrations of chloride \citep{sedriks1982corrosion,mangat1988corrosion,sridhar2004predicting,lv2015study,wu2017effect}. Due to prolonged exposure in such an aggressive atmospheric environment, these steels are susceptible to localized corrosion, which can lead to high maintenance costs as well as potential threats to structural health and safety \citep{ernst2007explanation,caines2015experimental,chen2015influence,schorr2016corrosion}. To mitigate these risks, it is necessary to fundamentally understand and predict the propagation of localized corrosion. Existing studies have shown that localized corrosion is a complex electrochemical dissolution process occurring at the interface between metal (solid) and electrolyte (liquid) \citep{ernst2002pitI,Frankel2008}. The stable propagation of localized corrosion is governed by the bulk environment, the composition of the alloy, and the aggressive chemistry within the localized corrosion site \citep{yin2017numerical,katona2019prediction}. To better understand the critical chemistry that enables or prevents the localized corrosion of 300 series stainless steels, we herein conduct numerical studies using the multi-species reactive transport model \citep{sun2019sequential} to investigate key electrochemical factors responsible for stable pitting and repassivation of 316L stainless steel wire inside the sodium chloride solution.

The stable propagation of an active pit requires an aggressive chemistry: high concentration of metal cations produced by the oxidation of the metal alloys and low (acidic) pH caused by the subsequent hydrolysis of those cations \citep{mankowski1975studies,strehblow1976electrochemical,bates1976hydrolysis}. The conditions that ensure the maintenance of such chemistry were mathematically described by Galvele through analysis of the steady state relationship between metal dissolution and mass transport inside a one-dimensional (1-D) pit \citep{galvele1976transport}. In that study, it was theoretically demonstrated that the product of the current density and the pit depth needed to be greater than a critical value to maintain the minimum local aggressive chemistry for continued stable pit growth.  This critical value has been referred to as the pit stability product ($X_\text{ps}$) \citep{pistorius1992metastable,burstein1993nucleation} and can be evaluated through artificial pit experiments \citep{srinivasan2015high}. There are several critical electrochemical factors involved in artificial pit experiments to identify the stable pit growth under a salt film, film-free metal dissolution and transition to pit repassivation \citep{woldemedhin2015effects,jun2020corrosion}. Typical experiments of this nature usually apply a high electrical potential at the beginning stage of experiments in order to initiate and grow pits, which leads to the precipitation of a salt film on the corroding surface \citep{isaacs1973behavior,tester1975diffusional}. The precipitation of a salt film further results in diffusion-limited metal dissolution so that the corroding system is in a quasi-steady state. The pit stability product under a salt film can then be extracted from measuring the diffusion-limited current density at different pit depths \citep{newman1983diffusion,gaudet1986mass,laycock1997localised,moayed2006relationship}. In order to estimate the conditions that govern the transition of the pit from stability to repassivation, rapid polarization scans can be employed which decrease the applied potential below the point where the salt film disappears. In this region, metal dissolution proceeds in a salt film-free stage. As the applied potential continues to be decreased, the corresponding decrease in metal dissolution rate and subsequently diminished cation hydrolysis \citep{mankowski1975studies} would lead to a rise of the pH near the corroding surface and eventually cause the pit to passivate. The repassivation potential is the potential corresponding to these conditions below which the pit will no longer propagate \citep{szklarska1978crevice,Anderko2004,anderko2008general,jafarzadeh2018peridynamic}.

Complementary to experimental studies of pitting corrosion, mechanistic modeling studies are essential to better understand the chemistry evolution of both active and passive pits. The simplest of mechanistic models consider mass transport of solute species (e.g., metal or hydrogen ions) in the solution environment based on the Fick's law of diffusion \citep{fick1855ueber}. Therefore, the 1-D Fickian diffusion model has been widely used to simulate the evolution of the chemistry inside the 1-D pit \citep{alkire1979location,laycock1998perforated,jun2015effect,srinivasan2016geometric,srinivasan2016evaluating,srinivasan2016one}. Srinivasan et al. used this model to evaluate the effect of the external hemispherical boundary layer on the cation flux inside the pit \citep{srinivasan2016geometric} and to demonstrate the dilution of cations near the corroding surface when repassivation happens \citep{srinivasan2016evaluating,srinivasan2016one}. Jun et al. applied the similar model to analyze the effects of chloride concentration and temperature on 1-D pit growth \citep{jun2015effect}. However, the main drawback of the 1-D diffusion model is that it ignores the effect of electro-migration and viscosity on species transport. Therefore, Jun et al. further extended their model to calculate the flux inside the pit by incorporating the effect of electro-migration and viscosity into the species diffusion coefficient based on the Stokes-Einstein equation \citep{jun2016further}. But this improvement is based only on parametric calibration which does not fundamentally correct the governing physics of species mass transport, and thus may not be applicable when focusing on the species concentration and the electrical potential instead of solely the flux. Furthermore, the evolution of chemistry dominated by chemical reactions occurring inside the pit usually cannot be modeled together with the 1-D diffusion model simultaneously and may require additional commercial software and databases. Consequently, it is ideal to use a comprehensive mechanistic model of pitting corrosion, which considers electro-diffusive mass transport of multiple species and chemical reactions in solution. In this study, we adopt a multi-species reactive transport model based on the sequential non-iterative approach (SNIA) developed by Sun and Duddu \citep{sun2019sequential}. Comparing to other numerical approaches proposed for solving electro-diffusive-reactive transport equations \citep{sharland1988mathematical,sharland1988mathematicalb,walton1990mathematical,walton1996numerical,chen2015peridynamic,mai2016phase,mai2018new}, the SNIA is able to establish the pit chemistry with multiple species and reactions in a more accurate and flexible manner.

The remainder of this paper is organized as follows: in \sref{sec:background}, we briefly introduce the setup of artificial pit experiments and two commonly used mass transport models for simulating multi-species reactive transport during corrosive dissolution; in \sref{sec:framework}, we establish the decoupled governing equations of the reactive-transport model for localized corrosion and summarize the strategy we employ to solve those equations; in \sref{sec:parameterization}, we detail the calibration of parameters as the input of the model based on experimental data; in \sref{sec:results}, we present the predicted critical electrochemical factors for both pits stably corrosion under a salt film and transitioning from film-free stable pitting to repassivation, and then discuss the potential mechanisms that dominate the evolution of the local chemistry for pit stability and repassivation; in \sref{sec:conclusion}, we conclude with a brief summary and closing remarks.

\section{Background} \label{sec:background}
In this section, we first introduce the setup of artificial pit experiments, from which we extract data to calibrate parameters of electrode kinetics. We next briefly summarize two commonly used mass transport models and their limitations for simulating multi-species reactive transport during corrosive dissolution.

\subsection{Artificial pit experiments} \label{sec:experiment}
\begin{figure}
  \centering
  \includegraphics[width=\textwidth,height=\textheight,keepaspectratio]{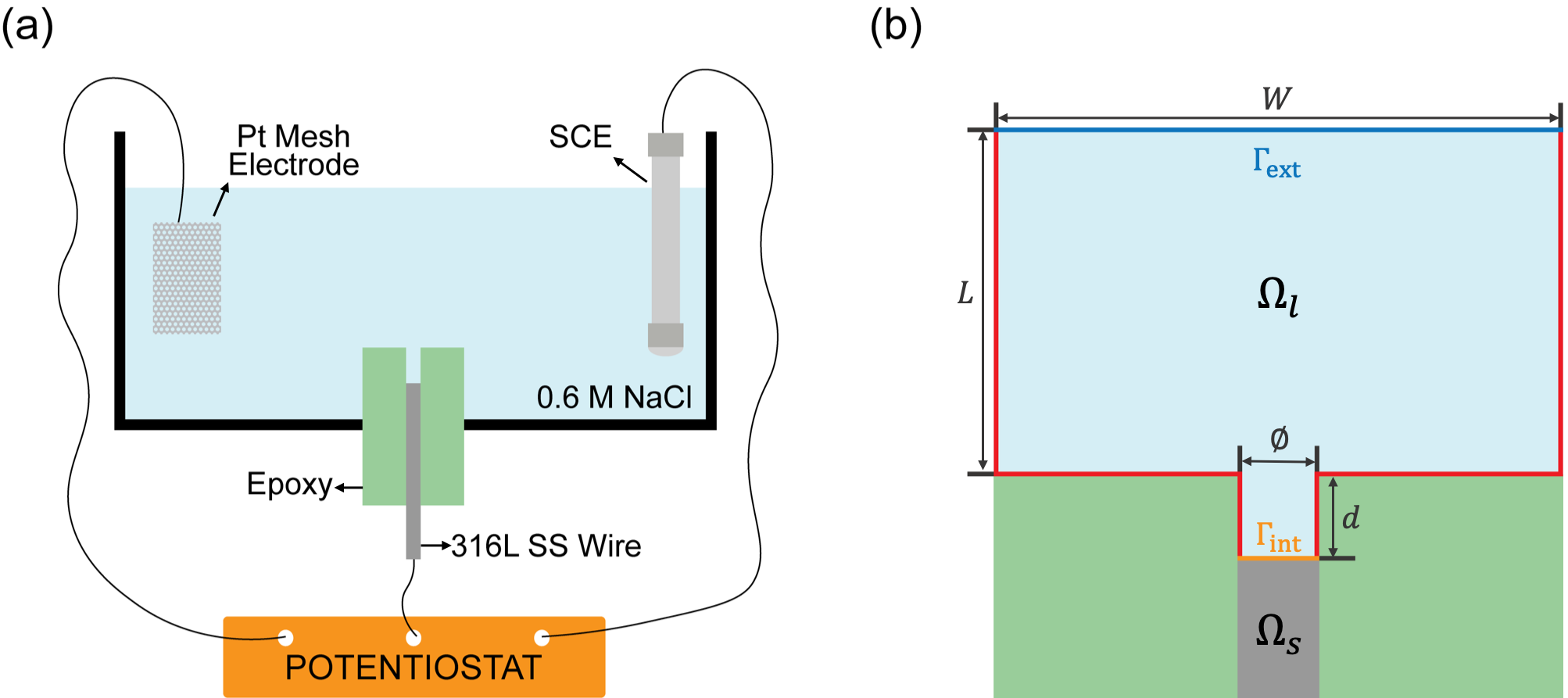}
  \caption{Schematic diagram of: (a) the experimental setup; (b) the two-dimensional domain for the artificial pit corrosion problem. The corrosion interface $\Gamma_{\text{int}}$ (orange) between solid metal $\Omega_s$ (gray) and liquid solution $\Omega_l$ (light blue) domains. Far-field (or bulk solution) concentrations and applied potential are prescribed at the external (dark blue) boundary $\Gamma_{\text{ext}}$. No flux condition is enforced on the boundaries marked by red color. The rest of the domain (green) is composed of Epoxy and is excluded from the simulation domain.} \label{fig:expset&domain}
\end{figure}

This study utilized data from one-dimensional artificial pit experiments reported by Srinivasan et al. \citep{srinivasan2015high,srinivasan2016one}. As shown in \fref{fig:expset&domain}(a), the artificial pit was constructed using 316L stainless steel wires with diameter $\diameter=50.8\ \mu$m embedded in epoxy. The wires were composed of $67.98$ wt\% \ce{Fe}, $17.07$ wt\% \ce{Cr}, $10.66$ wt\% \ce{Ni}, and $2.16$ wt\% \ce{Mo} with trace amounts of other elements, which were ignored in this study. After surface polishing to a finish of 320 grit with \ce{SiC} abrasive paper, the electrode was placed upright in the test container filled with $0.6$ M \ce{NaCl} solution. A saturated calomel electrode (SCE) and a platinum mesh electrode were employed as the reference and counter electrode, respectively. All tests were performed at an average ambient temperature of $22^{\circ}\mathrm{C}$ using a Bio-Logic SP-200 (Bio-Logic SAS, Claix, France) potentiostat. The pit was initiated with a potential of $+750\ $mV$_{\text{SCE}}$ and then propagated to different depths by applying a lower potential of $+450\ $mV$_{\text{SCE}}$. After that, a rapid cathodic polarization scan at 5 mV/s to a final potential of $-900\ $mV$_{\text{SCE}}$ was carried out to extract the pit stability product, the transition and the repassivation potentials for a specific pit geometry. Parametric calibration based on the experimental data will be detailed in Section \ref{sec:parameterization}.

\subsection{Existing mathematical models} \label{sec:existmodel}
Based on the geometric setup of artificial pit experiments, we define a two-dimensional domain that consists of two phases: the solid electrode domain $\Omega_s$ and the liquid electrolyte domain $\Omega_l$, as shown in \fref{fig:expset&domain}(b). The electrode and electrolyte domains are separated by a sharp corroded interface $\Gamma_{\text{int}}$. In the electrolyte domain $\Omega_l$, both mass transport and homogeneous chemical reactions can lead to the change of the species concentration. According to the law of conservation of mass, the rate of change of the concentration of the species $i$ can be expressed as 
\begin{equation}\label{eq:ratechangeofspecies}
	\frac{\partial C_i}{\partial t}=-\nabla\cdot\boldsymbol{J}_i+R_i~~\text{in}~~\Omega_l \text{,}~~i=\left\{1,2,\cdots,n\right\},
\end{equation}

\noindent where $C_i$ [mol/m$^3$] and $\boldsymbol{J}_i$ [mol/m$^2$/s] are the concentration and flux density of the $i^{\text{th}}$ species at time $t$ [s], respectively; $\nabla$ denotes the spatial gradient vector; $R_i$ [mol/m$^3$/s] is the rate of chemical reactions and $n$ denotes the total number of species in $\Omega_l$. In the electrolyte environment, the transport of species can be attributed to the effects of the species concentration gradient and/or the electrical potential gradient \citep{sharland1988mathematical}. Depending on whether the electro-migration effect is taken into consideration or not, the flux $\boldsymbol{J}_i$ can be described by two different models: Fickian diffusion model and Nernst-Planck model.

According to Fick's laws of diffusion \citep{fick1855ueber}, the magnitude of species flux density $\boldsymbol{J}_i$ is proportional to the concentration gradient, which can be written as
\begin{equation}\label{eq:fickianflux}
	\boldsymbol{J}_i=-D_i\nabla C_i,
\end{equation}

\noindent where $D_i$ [m$^2$/s] is the Fickian diffusion coefficient of species $i$. Furthermore, when there exist multiple ions in the electrolyte during corrosion, charge balance needs to be satisfied by enforcing the local electro-neutrality (LEN) condition
\begin{equation}
    \sum_{i=1}^{n}z_iC_i=0~~\text{in}~~\Omega_l \label{eq:LEN}.
\end{equation}

\noindent where $z_i$ [--] is the charge number of species $i$. Therefore, there will be $n+1$ equations in total with only $n$ unknowns as the concentration field of each species. Under such circumstances, the system is overdetermined and does not have a unique solution.

The Nernst-Planck model extends the Fick's law of diffusion by considering the transport of species due to both ionic diffusion and electro-migration. By assuming the electrolyte to be in a stagnant condition, advection can be ignored, and the flux $\boldsymbol{J}_i$ of the $i^{\text{th}}$ species can be written as
\begin{equation}\label{eq:npflux}
	\boldsymbol{J}_i=-D_i\nabla C_i-\frac{z_iF}{R^*T}D_iC_i\nabla\psi,
\end{equation}

\noindent where $F$ [C/mol] is the Faraday's constant, $R^*$ [J/mol/K] is the universal gas constant, $T$ [K] is the temperature, and $\psi$ [V] is the electrical potential of electrolyte. The electrical potential can be established by solving the LEN Equation \eref{eq:LEN}, which makes the system well-defined and has a unique solution. However, artificial pit experiments demonstrated that the electrolyte near the electrode will reach to saturation when being applied high enough electrical potential \citep{srinivasan2016one}. Because the Nernst-Planck model is used to describe the mass transport of aqueous chemical species in dilute electrolytic solutions \citep{sharland1988mathematical,kontturi2008ionic}, we need to further enrich Equation \eref{eq:npflux} by considering the effect of viscosity associated with temperature on the diffusivity of the species in the highly concentrated electrolytic environment.

\section{Numerical Framework} \label{sec:framework}
In this section, we first detail the decoupled multi-species reactive transport equations corresponding to the localized corrosion problem. In order to solve those equations in an efficient and accurate manner, we next briefly summarize the employed solution strategy based on the sequential non-iterative approach.

\subsection{Decoupled formulation} \label{sec:decoupleformulation}
According to the sequential non-iterative approach (SNIA) \citep{arnold2017influence,sun2019sequential}, we split Equation \eref{eq:ratechangeofspecies} into two separate equations describing mass transport and chemical reaction in strong ionic solutions, respectively. The rate of chemical reaction $R_i$ will be first eliminated from the species transport Equation \eref{eq:ratechangeofspecies}, which will yield physically incorrect ionic concentrations. The so-called ``uncorrected'' concentrations $\widetilde{C}_i$ can be computed from
\begin{equation} \label{eq:masstransport}
    \frac{\partial\widetilde{C}_i}{\partial t}=D_i\nabla^2\widetilde{C}_i+\frac{z_iD_iF}{R^*T}\nabla\cdot\left(\widetilde{C}_i\nabla\psi \right)~~\text{in}~~\Omega_l \text{,}~~i=\left\{1,2,\cdots,n\right\}.
\end{equation}

Because the neglected chemical reactions satisfy charge balance, the LEN condition Equation \eref{eq:LEN} is still valid for the uncorrected concentrations and can be enforced to establish the electrical potential field. Thus, the uncorrected concentrations and electrical potential can be determined by solving Equation \eref{eq:masstransport} along with LEN condition Equation \eref{eq:LEN}. The boundary conditions corresponding to Equations \eref{eq:LEN} and \eref{eq:masstransport} are given by
\begin{equation} \label{eq:boundarycondition} 
\begin{aligned}
  \widetilde{C}_i(\textbf{x},t)~~=~~C_i^\infty~~\text{and}~~~\psi(\textbf{x},t) = -E_{\text{app}}~~~~&\text{on}~~\Gamma_{\text{ext}}, \\
  \boldsymbol{J}_i\cdot\textbf{n}=\frac{\kappa_ii_\text{a}}{z_iF}~~~~&\text{on}~~\Gamma_{\text{int}},
\end{aligned}
\end{equation}

\noindent where ${C}_i^\infty$ is the far-field concentration of species $i$, $E_{\text{app}}$ is the applied electrical potential, $\textbf{n}$ denotes the unit normal to the interface $\Gamma_{\text{int}}$ pointing outward from $\Omega_l$, and $i_\text{a}$ [A/m$^2$] represents the anodic current density. For metal cations dissolved into the electrolyte through anodic reactions, $\kappa_i$ [--] is the molar fraction of the corresponding metal element in the alloy; For hydroxide ion, $\kappa_i$ is the ratio of the local cathodic current density to the anodic current density.

One of the assumptions of the SNIA is that, characteristic times of chemical reactions in aqueous solution are much shorter than those of the mass transport or localized corrosion processes \citep{sharland1992mathematical,walton1996numerical}. Therefore, the concentration of species involved in chemical reactions is always satisfied with the chemical equilibrium condition as given by
\begin{equation} \label{eq:chemicalequilibrium}
	K_r\prod_{p}C_p^{\nu_{rp}}=\prod_{q}C_q^{\nu_{rq}}.
\end{equation}

\noindent where $K_{r}$ is the equilibrium constant of the $r^{\text{th}}$ reaction, $C_p$ and $C_q$ are concentrations of the reactant and product species, respectively, and $\nu_{rp}$ denotes the stoichiometric coefficient of species $p$ in the $r^{\text{th}}$ reaction. The corrected concentrations $C_i$ that satisfy the chemical equilibrium condition described above are related to the uncorrected concentrations $\widetilde{C}_i$ as \citep{sun2019sequential}
\begin{equation} \label{eq:correctedconcentration}
	C_i=\widetilde{C}_i+\sum_{r=1}^{n_r}\omega_{ri} \nu_{ri} \widehat{C}_r,
\end{equation}

\noindent where $n_r$ is the total number of chemical reactions occurring in the electrolyte, the product of $\nu_{ri}$ and $\widehat{C}_r$ is the absolute value of concentration change of species $i$ due to the $r^{\text{th}}$ chemical reaction, and $\omega_{ri}$ represents whether species $i$ is a reactant, product or not involved in the forward direction of the $r^{\text{th}}$ reaction. If the species $i$ is not involved in the $r^{\text{th}}$ reaction, $\omega_{ri}=0$; however, if involved in the forward reaction as the product then $\omega_{ri}=1$, or as the reactant then $\omega_{ri}=-1$. Substituting Equation \eref{eq:correctedconcentration} into Equation \eref{eq:chemicalequilibrium}, we can obtain $n_r$ number of nonlinear algebraic (chemical reaction) equations to solve for unknowns $\widehat{C}_r$ as follows:
\begin{equation} \label{eq:chemicalequation}
	K_r\prod_{p}\left(\widetilde{C}_p+\sum_{l=1}^{n_r}\omega_{lp}\nu_{lp}\widehat{C}_l\right)^{\nu_{rp}}-\prod_{q}\left(\widetilde{C}_q+\sum_{l=1}^{n_r}\omega_{lq}\nu_{lq}\widehat{C}_l\right)^{\nu_{rq}}=0\text{,}~~r=\left\{1,2,\cdots, n_r \right\}.
\end{equation}

\noindent To ensure the non-negativity of species concentrations, the nonlinear chemical reactions equations \eref{eq:chemicalequation} are solved using the Newton-Raphson method together with the under-relaxation technique \citep{sun2019sequential}.

\subsection{Solution strategy} \label{sec:solstrategy}
Based on the SNIA \citep{sun2019sequential}, we employ a two-step staggered numerical solution strategy at any given time increment as follows:
\begin{enumerate}
    \item Compute the uncorrected concentrations $\widetilde{C}_i$ and electrical potential $\psi$ from the Nernst-Planck equations \eref{eq:masstransport} and the LEN condition \eref{eq:LEN} along with boundary conditions defined in Equation \eref{eq:boundarycondition};
    \item Compute the corrected concentrations $C_i$ of each species by solving the chemical equilibrium equations \eref{eq:chemicalequation} using the modified Newton-Raphson method.
\end{enumerate}

\section{Simulation and Calibration} \label{sec:parameterization}
In this section, we first clarify the two major stages involved in the numerical study: stable pitting under a salt film and film-free dissolution that eventually transitions to repassivation. We then detail the calibration process for several key parameters including the elapsed time of each stage, kinetics of electrode and chemical reactions, diffusivity and viscosity of each species.

\subsection{Stages of pitting} \label{sec:stagetime}
We herein consider two main stages of the artificial pit experiments as the objectives of the numerical investigation: I. Stable dissolution under a salt film, and II. Film-free dissolution and transition to repassivation. In Stage I, the application of high electrical potential leads to metal dissolution and hydrolysis, which further results in a local aggressive chemistry inside the pit. Due to the continuous metal dissolution, a salt film precipitates on the corroding surface within the saturated metal chloride solution. The presence of the salt film results in diffusion-limited dissolution because it fixes the concentration at the bottom of the pit. The pit depth is then the diffusion distance. The pit depth can be calculated directly from the charge density passed based on the Faraday’s law \citep{srinivasan2015high}. Once the pit has been corroded to the expected depth, decreasing the applied potential dissolves the salt film and film-free dissolution ensues. As the applied potential is further reduced, the anodic dissolution rate and the associated current density across the electrode-electrolyte interface $\Gamma_{\text{int}}$ also decreases in response. The dilution of the aggressive chemistry at the pit base that follows due to a lower metal dissolution rate as well as diffusion of concentrated aggressive species out of the pit eventually leads to repassivation.

\begin{figure}
  \centering
  \includegraphics[width=\textwidth,height=\textheight,keepaspectratio]{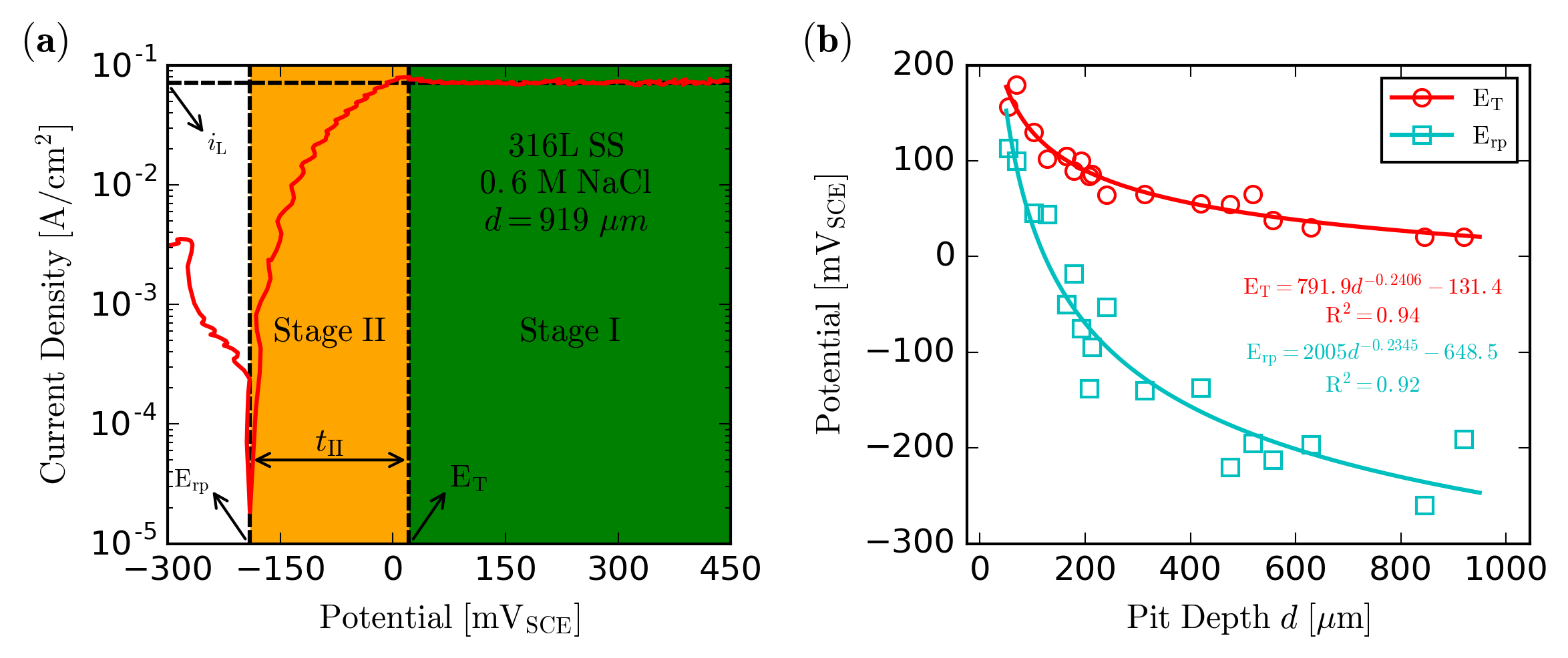}
  \caption{Extraction of kinetic parameters from experiments: (a) Polarization kinetics data indicating measurements of the diffusion-limited dissolution current density $i_\text{L}$, the transition potential $E_\text{T}$, and the repassivation potential $E_\text{rp}$. Red line represents the polarization curve obtained from the experiment. Green and orange regions are the first (stable pitting under a salt film) and second (film-free dissolution and transition to repassivation) stages of the study, respectively; (b) Estimation of the transition potential $E_\text{T}$ and the repassivation potential $E_\text{rp}$ with respect to the pit depth $d$ by least squares curve fitting based on the power law.} \label{fig:stagetime}
\end{figure}

The boundary between Stage I and Stage II can be identified using the experimental polarization kinetics data as plotted in \fref{fig:stagetime}(a). As long as the current density passing $\Gamma_{\text{int}}$ stays at the diffusion-limited current density $i_\text{L}$, the system is in Stage I and the pit is expected to grow stably under a salt film. For this stage, we aim to investigate the pit chemistry at the saturated state. Therefore, we neglect the motion of the corroding interface so that the experimental elapsed time of Stage I is not applicable for numerical modeling. Instead, we directly run the simulation with the expected pit depth till the saturated state is attained. When the rapid polarization scan starts, we consider the beginning of Stage II as the time when the applied electrical potential decreases to the transition potential $E_\text{T}$, and the current density becomes less than $i_\text{L}$. At the end of this stage, the repassivation of the pit is estimated to occur when the current density is reduced to $i_\text{rp}=$30 $\mu$A/cm$^2$. The applied potential recorded instantly when the current density reaches $i_\text{rp}$ is referred to as the repassivation potential $E_\text{rp}$. Because the polarization scan rate $\dot{E}_\text{app}$ [V/s] is a constant, the elapsed time of the second stage $t_\text{II}$ can be estimated by

\begin{equation} \label{eq:secondstagetime}
	t_\text{II}=\frac{E_\text{T}-E_\text{rp}}{\dot{E}_\text{app}}.
\end{equation}

\noindent In order to evaluate the elapsed time of the second stage for pits with intermediate depths that are not captured in the experimental results, we further perform least squares curve fitting based on the power law for both the transition and repassivation potentials with respect to the pit depth ($d$), as shown in \fref{fig:stagetime}(b). 

\subsection{Electrode reactions and kinetics} \label{sec:electrode}
We consider three types of electrode reactions happening during the experiment: anodic metal dissolution, internal and external cathodic reactions \citep{srinivasan2016evaluating}. Anodic dissolution and the internal cathodic reaction occur on the corroding surface of the 316L stainless steel wire, which is directly exposed to sodium chloride solution. Anodic dissolution includes the oxidation of iron, chromium, nickel, and molybdenum:
\begin{eqnarray}
    \ce{Fe &\rightarrow& Fe^{2+}}\ce{+2e^{-}}, \label{eq:anodicfe} \nonumber\\
    \ce{Cr &\rightarrow& Cr^{3+}}\ce{+3e^{-}}, \label{eq:anodiccr} \nonumber\\
    \ce{Ni &\rightarrow& Ni^{2+}}\ce{+2e^{-}}, \label{eq:anodicni} \nonumber\\
    \ce{Mo &\rightarrow& Mo^{3+}}\ce{+3e^{-}}. \label{eq:anodicmo} \nonumber
\end{eqnarray}

\noindent The local cathodic reaction is considered to be the hydrogen evolution reaction (HER) which can proceed as the reduction of water or protons listed below, as the conditions inside the pit are oxygen-poor.

\begin{equation} \label{eq:cathodicreaction}
	\ce{2H_{2}O + 2e^{-} \rightarrow H_2 + 2OH^{-}}. \nonumber
\end{equation}

\noindent In the experiment described, the vast majority of the cathodic reaction normally occurs on the counter electrode. Considering the fact that the counter electrode is spatially distant from the pit, the external cathodic reaction is assumed not to affect the local chemistry inside the pit and is consequently ignored for the purpose of this study.

In order to fulfill the electrode flux boundary condition defined in Equation \eref{eq:boundarycondition}, we need to calibrate the anodic current density $i_\text{a}$ based on experimental results. In artificial pit experiments, the net current density $i_\text{net}$ can be directly monitored using the potentiostat. Theoretically, the net current density is a combination of both anodic current density ($i_\text{a}$) and local cathodic current density ($i_\text{c,local}$), which can be written as

\begin{equation} \label{eq:netcurrent}
	i_\text{net}=i_\text{a}-i_\text{c,local}.
\end{equation}

\noindent At high rates of anodic dissolution, the local cathodic reaction (HER) is much lower due to the anodic polarization \citep{srinivasan2016evaluating}. Therefore, the anodic current density is approximately equal to the measured net (diffusion-limited) current density ($i_\text{a}=i_\text{net}=i_\text{L}$), and the local cathodic reaction is neglected during Stage I. However, once Stage II begins, the local cathodic reaction rate is considered to increase in accordance with the decrease of the anodic dissolution rate. Srinivasan and Kelly estimated the ratio of the local cathodic current density to the anodic current density to be about $0.3\%$ when the pit begins to repassivate \citep{srinivasan2016evaluating}. Based on this estimation, the anodic current density can then be calculated as

\begin{equation} \label{eq:anodiccurrent}
	i_\text{a}=\frac{i_\text{net}}{1-0.003}=\frac{i_\text{rp}}{1-0.003}.
\end{equation}

Regardless of the pit depth, we set the threshold of net current density to be a constant value of 30 $\mu$A/cm$^2$ in order to determine when repassivation occurs in artificial pit experiments. However, the diffusion-limited current density does vary for pits with different depths. Experiments have demonstrated that the product of the diffusion-limited current density $i_\text{L}$ and the pit depth $d$ is constant as long as the temperature and the bulk concentration of sodium chloride do not change \citep{srinivasan2015high}. This is the pit stability product $X_{\text{ps}}=i_\text{L}\cdot d$, which can be obtained from the linear regression analysis of the experimental data. Appendix \ref{appendix:pitstability} lists these values for stainless steel in \ce{NaCl} solution at different temperatures and bulk concentrations. 

\subsection{Diffusivity and viscosity}\label{sec:diffusivity}
The metal chloride solution near the corroding surface will reach saturation when a high enough electrical potential is applied. As demonstrated by Jun et al. \citep{jun2016further}, the effect of electrolyte viscosity on mass transport in such a concentrated electrolyte cannot be ignored. The effect of viscosity $\eta$ and temperature $T$ on the diffusion coefficient can be described using the Stokes-Einstein equation

\begin{equation} \label{eq:stokeseinstein}
	D=\frac{kT}{6\pi\eta r},
\end{equation}

\noindent where $k$ [J/mol/K] is the Boltzmann constant and $r$ [m] is the radius of the species. For the sake of simplicity, we assume that the the effect of temperature on the radius of the species can be neglected. Therefore, the diffusion coefficient can be approximately calculated using its reference value $D_\text{ref}$ as

\begin{equation} \label{eq:diffusionref}
	D=\frac{TD_\text{ref}}{\eta T_\text{ref}},
\end{equation}

\noindent where the reference temperature $T_\text{ref}$ is taken as the room temperature $298.15$ K. The reference values of diffusion coefficients for all species considered in this study have been listed in Appendix \ref{appendix:diffusivity}. The viscosity of the electrolyte can be analytically represented based on the extended Jones--Dole equation \citep{aleksandrov2012viscosity}, which can be written as
\begin{equation} \label{eq:viscosity}
	\eta=1+a_0{C_l}^\frac{1}{2}+a_1{C_l}+a_2{C_l}^2+a_3{C_l}^3,
\end{equation}

\noindent where $C_l$ is the concentration of the electrolyte, and coefficients $a_i$ can be calculated from its temperature-pressure dependence
\begin{equation} \label{eq:viscosityco}
	a_i=\sum_{m=0}^1\sum_{n=0}^4\zeta_{mn}p^m\tau^n,~~i=\left\{0,1,2,3\right\},
\end{equation}

\noindent where $p$ is the pressure, $\tau=T_0/T=400/T$ is the relative temperature, and $\zeta_{mn}$ is the coefficient calibrated from a series of experiments \citep{aleksandrov2012viscosity}, which has been listed in Appendix \ref{appendix:viscosity}. As noted by Aleksandrov et al., Equation \eref{eq:viscosity} is valid for describing the viscosity of aqueous solutions of sodium chloride in the concentration ranging from 0 to 6 M at temperatures from 0 to 325$^{\circ}\mathrm{C}$. Based on the assumption of the LEN condition \eref{eq:LEN}, we herein use the concentration of \ce{Cl^-} as the estimation of the electrolyte concentration $C_l$ to calculate the viscosity of each species.

\subsection{Chemical reactions and kinetics}\label{sec:chemicalreaction}
To capture the evolution of local chemistry inside the pit, we consider three types of chemical reactions taking place in the electrolyte phase $\Omega_l$ including water dissociation, hydrolysis and chloride complexation of metal ions, as follows
\begin{eqnarray}
    \ce{H^{+} + OH^{-} &\rightleftharpoons& H_{2}O}, \label{eq:waterdissociation} \nonumber\\
    \ce{M^{n+} + aH_{2}O &\rightleftharpoons& MOH^{(n-a)+}_a + aH^{+}}, \label{eq:metalhydrolysis} \nonumber\\
    \ce{M^{n+} + aCl^- &\rightleftharpoons& MCl^{(n-a)+}_a}. \label{eq:metalclcomplx} \nonumber
\end{eqnarray}

\noindent where \ce{M} refers to the constituent metallic elements (iron, chromium, and nickel) that dissolve into electrolyte as cations through the oxidation reaction. For ferrous and nickel ions, $n=2$ and $a=1$ or $2$; For chromium and molybdenum ions, $n=3$ and $a=1$, $2$ or $3$. The calculation of equilibrium constants $K_r$ for all three types of reactions has been summarized in Appendix \ref{appendix:reaction}.

\section{Results and Discussion} \label{sec:results}
In this section, we consider a 2-D model of pitting corrosion as depicted in \fref{fig:expset&domain}(b). We assume a constant pit diameter $\diameter=50\ \mu$m and different pit lengths $d=\{50,$ $100,$ $150,$ $200,$ $250,$ $350,$ $500,$ $750,$ $1000\}\ \mu$m. Because the morphology evolution of the corroding surface is neglected, the geometry of the pit does not change while dissolution is taking place. Following the example in \citep{srinivasan2016geometric}, we take the dimension of the rectangular bulk electrolyte region above the pit to be $W\times L=20\times10\ \text{cm}^2$. As suggested by Sun and Duddu \citep{sun2019sequential}, the size of time increment $\Delta t$ should be less than $1\%$ of the characteristic diffusion time scale $t_d = d^2/4D$ to ensure sufficient accuracy of the sequential non-iterative approach. Therefore, we herein take the size of time increment $\Delta t=0.01$ second. The problems are solved using the standard finite element method in the open-source software FEniCS \citep{alnaes2015fenics}. Based on artificial pit experiments, the remainder of this section is organized by two experimental stages: stable pitting and repassivation.

\subsection{Stage I: Stable pitting under a salt film} \label{stageI}
At the first stage, we conduct a series of numerical studies by varying the pit depth, the temperature and the bulk concentration of the electrolyte. The initial conditions of concentrations and the electrical potential fields can be expressed as

\begin{equation} \label{eq:initialstageI}
	\widetilde{C}_i(\textbf{x},0)~~=~~C_i^\infty~~\text{and}~~\psi(\textbf{x},0) = 0~~~~\text{in}~~\Omega_{l}.
\end{equation}

\noindent We herein define the metal cation ``Me'' as all species inside the pit that containing the primary alloying elements of 316L stainless steel (i.e., \ce{Fe}, \ce{Cr}, \ce{Ni}, \ce{Mo}). From in situ X-ray studies on solutions of 18-8 stainless steel in chloride media conducted by Isaacs et al. \citep{isaacs1995situ}, the saturated concentration of the metal cation $C_\text{Me}$ near the pit base is reported as $C_\text{sat}=5.02$ M. Therefore, simulations are run till the metal cation at the pit base reaches to saturation (i.e., $C_\text{Me}=5.02$ M). 

\subsubsection{Local chemistry}

\begin{figure}
  \centering
  \includegraphics[width=\textwidth,height=\textheight,keepaspectratio]{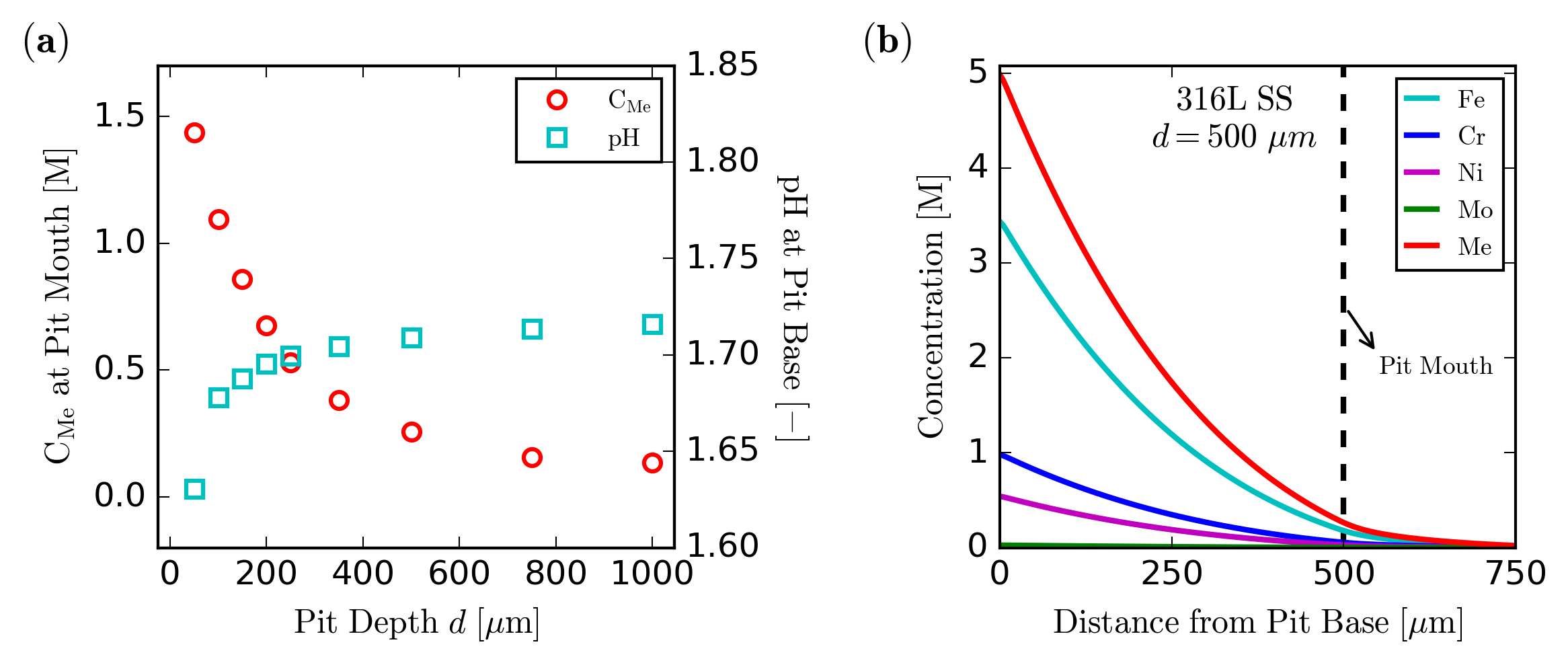}
  \caption{Numerical results of local chemistry inside the pit at the saturated state for pitting corrosion of 316L stainless steel wire within $0.6$ M \ce{NaCl} solution under room temperature $T=25^{\circ}\mathrm{C}$: (a) The concentration of metal cation at the pit mouth and the pH at the pit base with respect to different pit depths; (b) Concentration profiles of cations inside the pit with depth $d=500\ \mu$m.} \label{fig:stage1-candph}
\end{figure}

We first investigate the local chemistry inside the pit in the saturated state at different depths. The temperature $T$ and the bulk concentration of the sodium chloride electrolyte $C^\infty$ are taken as $25^{\circ}\mathrm{C}$ and $0.6$ M, respectively. In \fref{fig:stage1-candph}(a), we show the concentrations of metal cation at the pit mouth together with the pH at the pit base with respect to different pit depths. The concentration of metal cation at the pit mouth decreases from $1.4$ M ($\sim28\%C_\text{sat}$) to $0.2$ M ($\sim4\%C_\text{sat}$) associated with the increase of the pit depth from $50$ to $1000\ \mu$m. This trend has also been captured by the numerical results obtained from the 1-D Fick's diffusion model \citep{srinivasan2016geometric}. For shallow pits, the characteristic length of cations mass transport is longer than the pit depth. Therefore, there exists the external hemispherical boundary layer such that the concentration of the metal cation outside the pit does not immediately vanish at the pit mouth. As presented by Srinivasan et al. \citep{srinivasan2016geometric}, the flux of metal cations $J_\text{Me}$ inside the pit can be calculated as

\begin{equation} \label{eq:1ddiffusionflux}
	J_\text{Me}=\frac{D\Delta C_\text{Me}}{d}=\frac{D(C_\text{Me}^\text{base}-C_\text{Me}^\text{mouth})}{d}=\frac{DC_\text{sat}}{d}.
\end{equation}

\noindent Note that this calculation of flux is only based on the 1-D Fick's law of diffusion. In order to ensure that $C_\text{Me}$ at the pit mouth approaches zero so that accurate results can be obtained by this calculation, Srinivasan et al. \citep{srinivasan2016geometric} suggest choosing pit depths that are $8$ to $10$ times the pit diameter. Our predicted results also indicate a similar conclusion: for pits with diameter $\diameter=50\ \mu$m, $C_\text{Me}$ at the pit mouth is negligible ($<4\%C_\text{sat}$) only when the pit depth reaches $500\ \mu$m ($10\diameter$). However, owing to the effect of electrolyte viscosity and electro-migration, the flux cannot be calculated using Equation \eref{eq:1ddiffusionflux} for the numerical framework in our study. As shown in \fref{fig:stage1-candph}(b), the linear relationship between the concentration of cations and the spatial coordinate governed by Equation \eref{eq:fickianflux} is no longer valid. The flux may vary non-linearly inside the pit and can only be calculated using Equation \eref{eq:npflux}.

From \fref{fig:stage1-candph}(a), we also find that the pH at the pit base slightly increases with pit depth. The predicted values of the pH are all within the range of $1.63$ to $1.71$, which represent a locally acidic environment near the corroding surface. Srinivasan and Kelly modeled the local chemistry near the corroding pit surface using the solution thermodynamics database of the OLI Analyzer Studio 9.2 (OLI Systems, Inc., Cedar Knolls, NJ) software \citep{srinivasan2016evaluating}. They first evaluated the oxide speciation as a function of pH and found that the lowest pH that enable the formation of metal oxide (\ce{CrO(OH)}), which isolate the pit base from acidic electrolyte, is about $2.65$. Based on this estimation, our predicted chemistry near the pit base is aggressive enough that repassivation would not occur in this stage. Srinivasan and Kelly \citep{srinivasan2016evaluating} also estimated the local pH near the pit base if only anodic dissolution was considered. This case corresponded to the absence of any local cathodic reaction, thus modeling the conditions inside the pit during high-rate anodic dissolution, as expected when a salt film is present. The resulting pH predictions of as low as $-0.25$ may be due to the fact that they ignored the mass transport of individual cation species, and uniformly saturated the mixed solution of \ce{FeCl_2}, \ce{CrCl_3}, \ce{NiCl_2}, and \ce{MoCl_3} in their calculations. The model introduced in this paper is more comprehensive with respect to individual cation transport and the predicted pH range at the corroding surface at high anodic dissolution ($1.63$ to $1.71$) shows good agreement with the experimental data reported by Oldfield and Sutton \citep{oldfield1978crevice}, which measured the pH values of highly concentrated mixed solutions of \ce{CrCl_3}$\cdot$6\ce{H_2O} and \ce{FeCl_2}$\cdot$4\ce{H_2O} to be in the the range of $1.25$ to $1.75$.

\subsubsection{Electrolytic resistance}

To further quantitatively validate our model, we next focus on the prediction of the specific resistance of the electrolyte. The electrolyte resistance in the pitting corrosion system consists of two parts: the resistance of the electrolyte in the pit, and that of the bulk solution \citep{gaudet1986mass}. Theoretically, the pit resistance $R_\text{p}$ is supposed to increase linearly with the increase of the pit depth, and the bulk resistance $R_\text{b}$ is a constant and does not depend on the size of the pit. Therefore, the electrolyte resistance $R_l$ is related to dimensions of the pit by
\begin{equation} \label{eq:electrolyteR}
    R_l=R_\text{b}+R_\text{p}=\alpha_r\left(\frac{1}{2\diameter}+\frac{4d}{\pi\diameter^2}\right),
\end{equation}

\noindent where $\alpha_r$ is the specific resistance of the electrolyte solution. The IR drop commonly refers to the decrease of electrical potential from the pit base to the mouth owing to the resistance of the electrolyte. According to Ohm's law, the IR drop within the pit can be expressed as
\begin{equation} \label{eq:ohmlaw}
    \Delta E_\text{IR}=iR_\text{p},
\end{equation}

\noindent where $i$ is the electric current that passes through the entire pitting system. At the first stage, the IR drop inside the pit can be calculated using the electrical potential at the pit base ($\psi_\text{b}$) and mouth ($\psi_\text{m}$) obtained from the SNIA. The corresponding current passing through the system is a constant that equal to the product of diffusion-limited current density ($i_\text{L}$) and section area of the 316L stainless steel wire. Therefore, the pit resistance can be estimated using the predicted electrical potential field obtained from the SNIA as
\begin{equation} \label{eq:SNIAresist}
    R_\text{p}=\frac{\Delta E_\text{IR}}{i}=\frac{\psi_\text{b}-\psi_\text{m}}{i_\text{L}\pi\left(\frac{\diameter}{2}\right)^2}.
\end{equation}

\noindent Substituting Equation \eref{eq:electrolyteR} into Equation \eref{eq:SNIAresist}, we can then calculate the specific resistance $\alpha_r$ of the pit based on the theoretical understanding of the electrolyte resistance as
\begin{equation} \label{eq:specresistance}
    \alpha_r=\frac{\psi_\text{b}-\psi_\text{m}}{i_\text{L}d}.
\end{equation}

\begin{figure}
  \centering
  \includegraphics[scale=0.9,keepaspectratio]{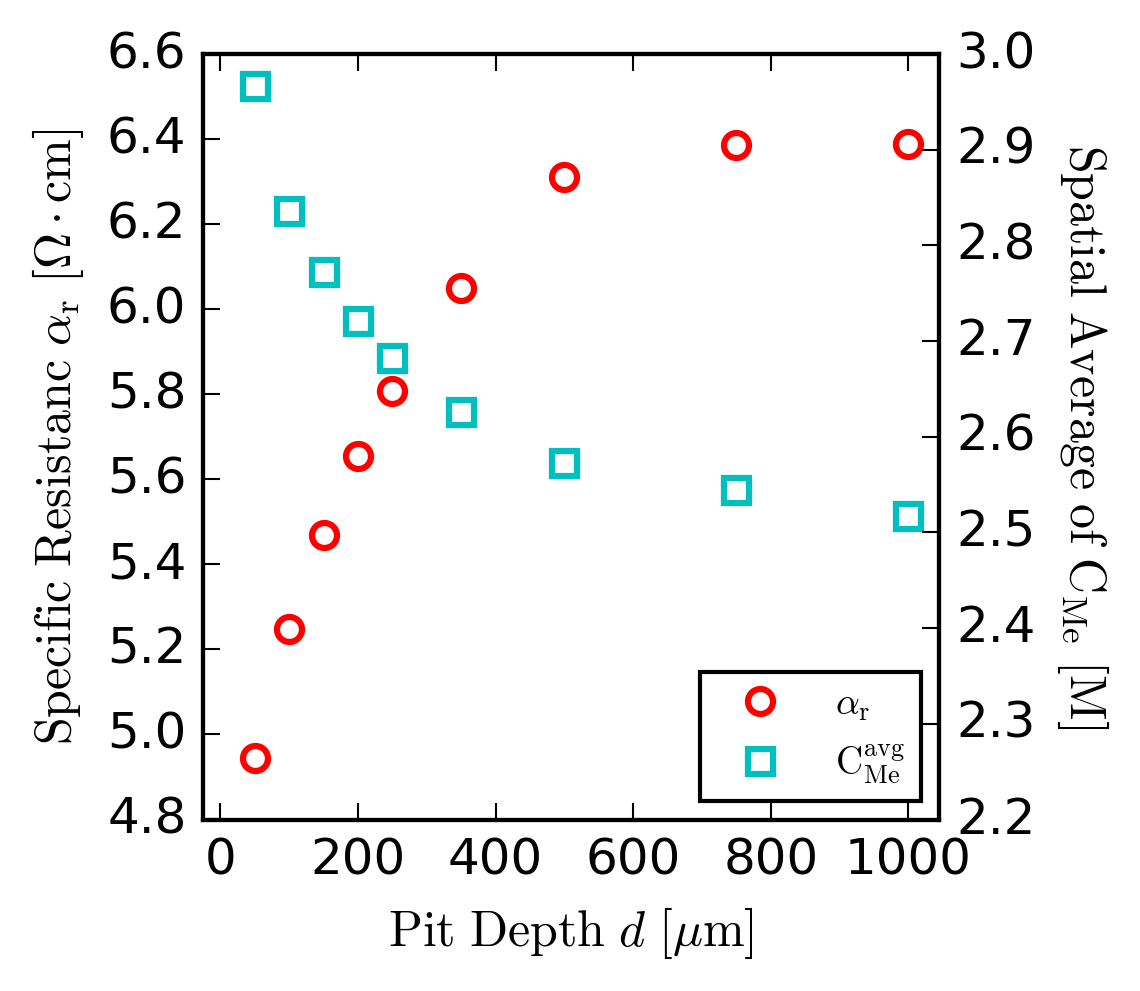}
  \caption{Numerical results of pit specific resistance and spatial average concentration of metal cation inside the pit with respect to different pit depths at the saturated state for pitting corrosion of 316L stainless steel wire within $0.6$ M \ce{NaCl} solution under room temperature $T=25^{\circ}\mathrm{C}$.} \label{fig:stage1-resist}
\end{figure}

Experiments have demonstrated that the specific resistance of solution decreases associated with the increase of solution concentration \citep{robinson2002electrolyte}. For the sodium chloride solution at room temperature $T=25^{\circ}\mathrm{C}$, the measured specific resistance decreases from $11.66$ to $4.368$ $\Omega\cdot$cm while the concentration of the solution increases from $1$ to $5$ M. In \fref{fig:stage1-resist}, we show predicted specific resistances of the pit electrolyte with respect to different pit depths. With the increase of the pit depth, the predicted specific resistance of the pit electrolyte increases from $4.9$ to $6.4$ $\Omega\cdot$cm. However, the concentration of the electrolyte is not uniformly distributed inside the pit, as shown in \fref{fig:stage1-candph}(b). To understand the relationship between the predicted $\alpha_r$ and the electrolyte concentration, we need to consider the average electrolyte concentration inside the pit. The spatially average concentration of metal cation inside the pit obtained from the SNIA is plotted in \fref{fig:stage1-resist}. It decreases from $3.0$ to $2.5$ M as the pit depth increase from $50$ to $1000\ \mu$m. We thus find that the specific resistance of the pit electrolyte and the concentration of the electrolyte are in inverse proportion, which is consistent with the experimental observation. Furthermore, the prediction obtained from our model shows a reasonable match with the experimental data in the quantitative level. Specifically, the predicted $\alpha_r$ and the spatially average $C_\text{Me}$ obtained from the SNIA are both included in the range of experimental results mentioned above.

\subsubsection{Effect of temperature and bulk concentration}

We next evaluate the effect of temperature and concentration of sodium chloride (bulk) solution on the local chemistry near the pit base and the specific resistance of the pit solution. We perform parametric studies on pitting corrosion of stainless steel by varying the temperature $T=\{25,$ $35,$ $45,$ $55\}\ ^{\circ}\mathrm{C}$ and bulk concentrations $C_\text{bulk}=\{0.6,$ $1.0,$ $3.0,$ $5.5\}\ \mathrm{M}$. To preclude any possible influence on those electrochemical factors caused by different pit depths, we here consider the pit with constant depth $d=500~\mu$m. All simulations are run until a final time $t^{\infty}=400~\text{seconds}$ to attain the saturated state of the solution inside the pit. 

\begin{figure}
  \centering
  \includegraphics[width=\textwidth,height=\textheight,keepaspectratio]{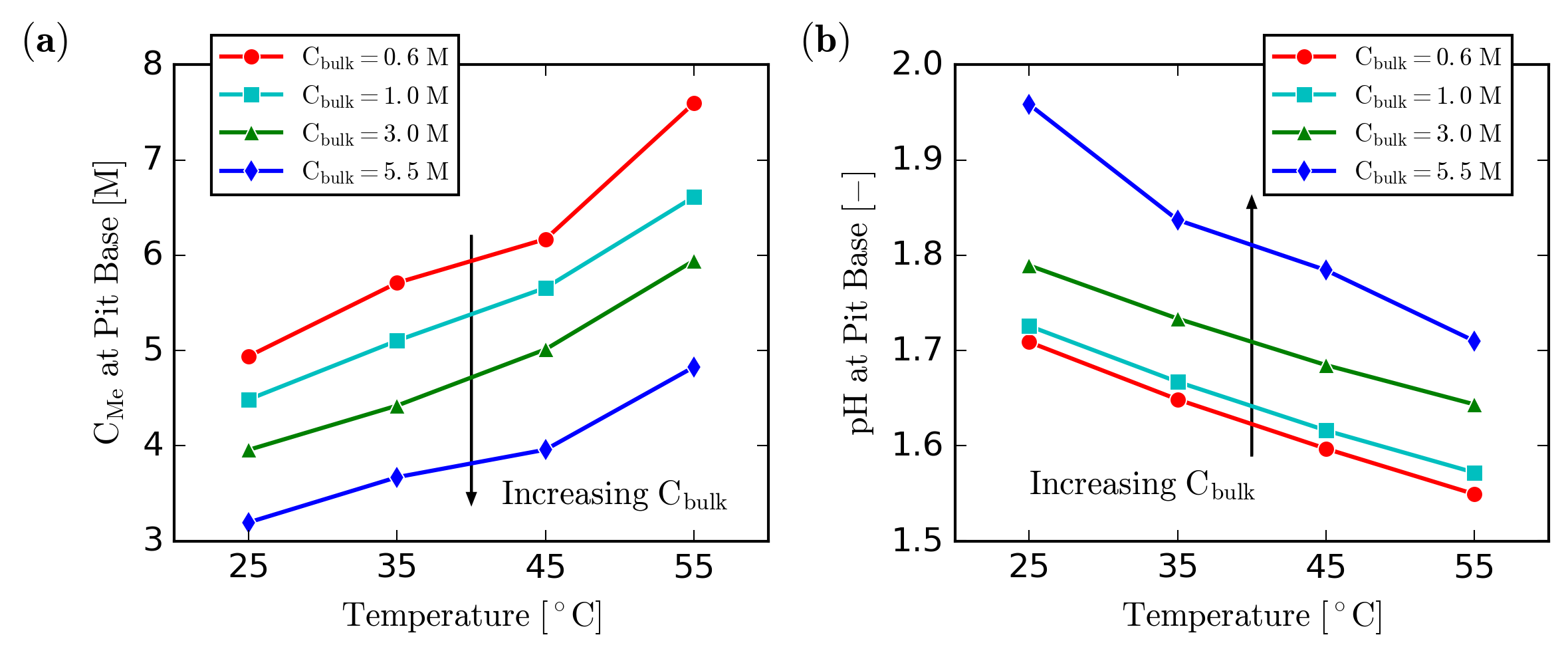}
  \caption{Parametric sensitivity of critical electrochemical factors with respect to temperature $T$ and bulk concentrations $C_\text{bulk}$: (a) Concentration of metal cation $C_\text{Me}$ at the pit base; (b) pH at the pit base.} \label{fig:stage1-tempcbulk}
\end{figure}

In \fref{fig:stage1-tempcbulk}, we demonstrate the effect of temperature and bulk concentration on the local chemistry near the pit base. It is evident that the saturation concentration of metal cation increases along with the increase of temperature, as shown in \fref{fig:stage1-tempcbulk}(a). The same trend has also been reported based on experimental investigation of the ferrous chloride solution \citep{chou1986solubility} and the corrosive electrolyte of 304L stainless steel \citep{ernst2002pitII,katona2019prediction}. This can be simply explained by the fact that the solubility of most salts increases with temperature \citep{pauling1988general}. From \fref{fig:stage1-tempcbulk}(a), we also find that the saturation concentration of the metal cation decreases as the bulk concentration increases. Because of the common ion effect, the solubility of the metal cation decreases significantly with an increase in chloride concentration. This inverse relationship between the saturation concentration of metal cation and the bulk concentration of sodium chloride has also been noted by several experimental \citep{ernst2007explanation} and numerical studies \citep{jun2015effect,jun2016further}. In \fref{fig:stage1-tempcbulk}(b), we show the dependence of pH at the pit base on the temperature and bulk concentrations. Based on our previous analysis, the metal cation at the pit base is more concentrated in the system at higher temperature and lower bulk concentration. Under such circumstances, the hydrolysis of metal cations that producing hydrogen ion is more intensive, which leads to the decrease of pH near the pit base \citep{bates1976hydrolysis,mankowski1975studies}. Therefore, the pH decreases associated with the increase of temperature and the decrease of the bulk concentration.

\begin{figure}
  \centering
  \includegraphics[scale=0.9,keepaspectratio]{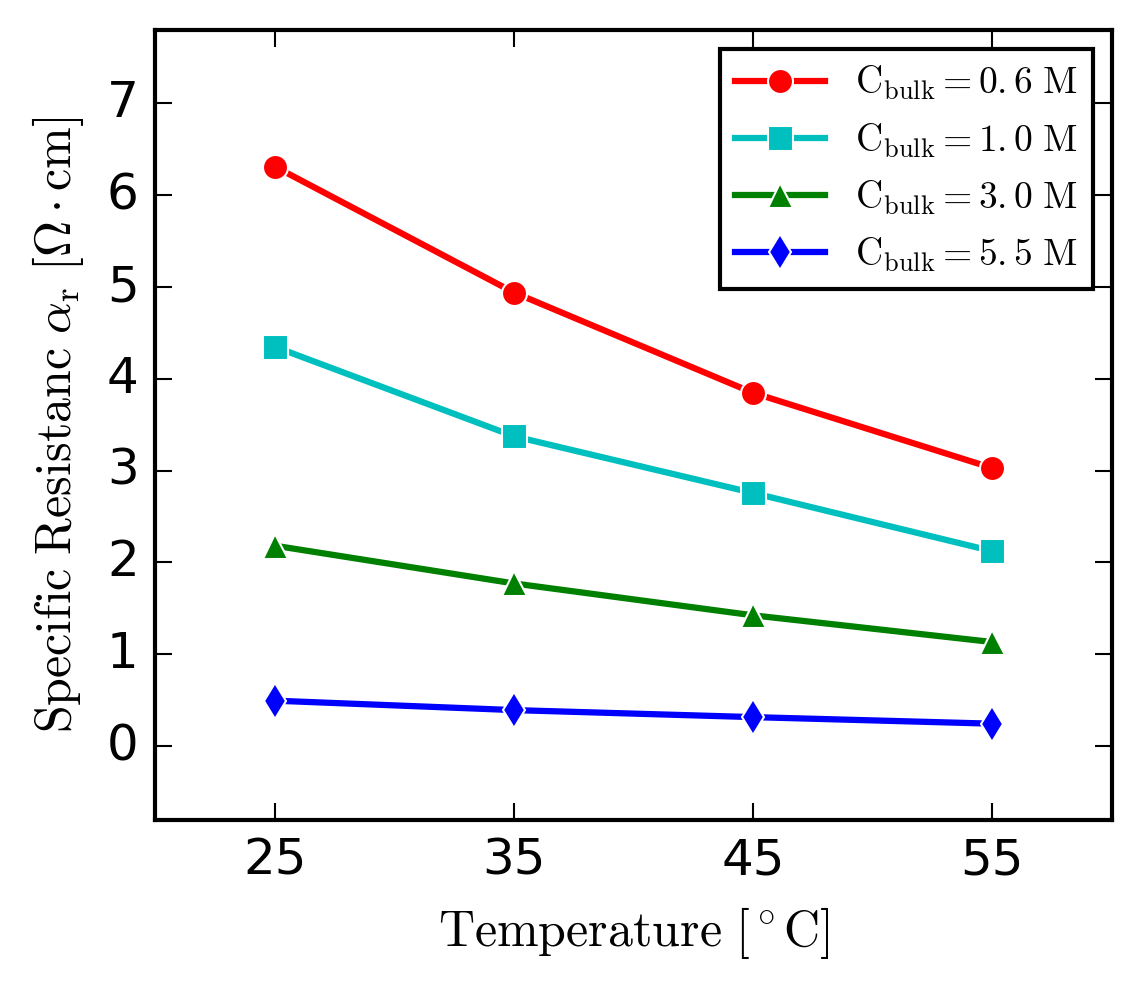}
  \caption{Parametric sensitivity of the pit specific resistance with respect to temperature $T$ and bulk concentrations $C_\text{bulk}$.} \label{fig:stage1-resistTC}
\end{figure}

In \fref{fig:stage1-resistTC}, we calculate the specific resistance of the pit solution with different temperature and bulk concentrations based on Equation \eref{eq:specresistance}. On the one hand, the pit specific resistance decreases associated with the increase of the temperature. This is due to the fact that the increase of the temperature can lead to two important changes of the electrolyte: Firstly, with the increase of the temperature, the interionic attraction decreases hence ionic mobility increases \citep{glasstone2011introduction}; Secondly, the degree of electrolyte dissociation increases together with the temperature so that the number of ions available for conduction increases \citep{bagotsky2005fundamentals}. On the other hand, the reduction of the pit specific resistance is expected when we consider more concentrated sodium chloride (bulk) solution. In general, this is valid because the electrolyte with higher concentration has more charged particles that are capable of carrying an electrical current. However, the positive correlation between the electrolyte concentration and conductivity may not always hold. As the electrolyte concentration increases above a certain value, the decrease of average distance between cation and anion can lead to more intensive interionic interaction and thereby limit the ionic mobility \citep{pitzer1973thermodynamics}. Such phenomena can be mathematically described using the activity coefficient (i.e., a function of the electrolyte ionic strength) \citep{stokes1948ionic,pitzer1973thermodynamics}, providing a potential direction for future model extension.

\subsection{Stage II: Film-free dissolution and transition to repassivation} \label{stageII}
At Stage II, we perform simulations of localized corrosion problems inside pits with different depths. The temperature and the concentration of the bulk solution are taken as the constant $T=25^{\circ}\mathrm{C}$ and $C_\text{bulk}=0.6$ M, respectively. All simulations are run until the final time $t_\text{II}$ calibrated from the experimental data as detailed in Section \ref{sec:stagetime} to ensure that the pit is repassivated. This stage is initialized by the conditions listed as follows

\begin{equation} \label{eq:initialstageII}
	\widetilde{C}_i(\textbf{x},0)~~=~~C_i^\text{sat}~~\text{and}~~\psi(\textbf{x},0) = \psi^\text{sat}~~~~\text{in}~~\Omega_{l}.
\end{equation}

\noindent where $C_i^\text{sat}$ and $\psi^\text{sat}$ are the saturated-state results of species concentrations and the electrical potential obtained from the first-stage simulation, respectively. As noted in Section \ref{sec:stagetime}, while the applied electrical potential is decreasing from the beginning of Stage II, the current density passing through the pit decreases accordingly. Therefore, transient flux boundary conditions calibrated from the experimental data are necessary to accurately model the evolution of chemistry near the pit base during repassivation. However, Srinivasan and Kelly \citep{srinivasan2016one} have shown that the critical pit chemistry determined using the transient flux boundary conditions is in good agreement with the value obtained using the constant flux boundary conditions. We thus apply the constant flux boundary condition based on the repassivation current density $i_\text{rp}$ as a reasonable simplification of the model.

\begin{figure}
  \centering
  \includegraphics[width=\textwidth,height=\textheight,keepaspectratio]{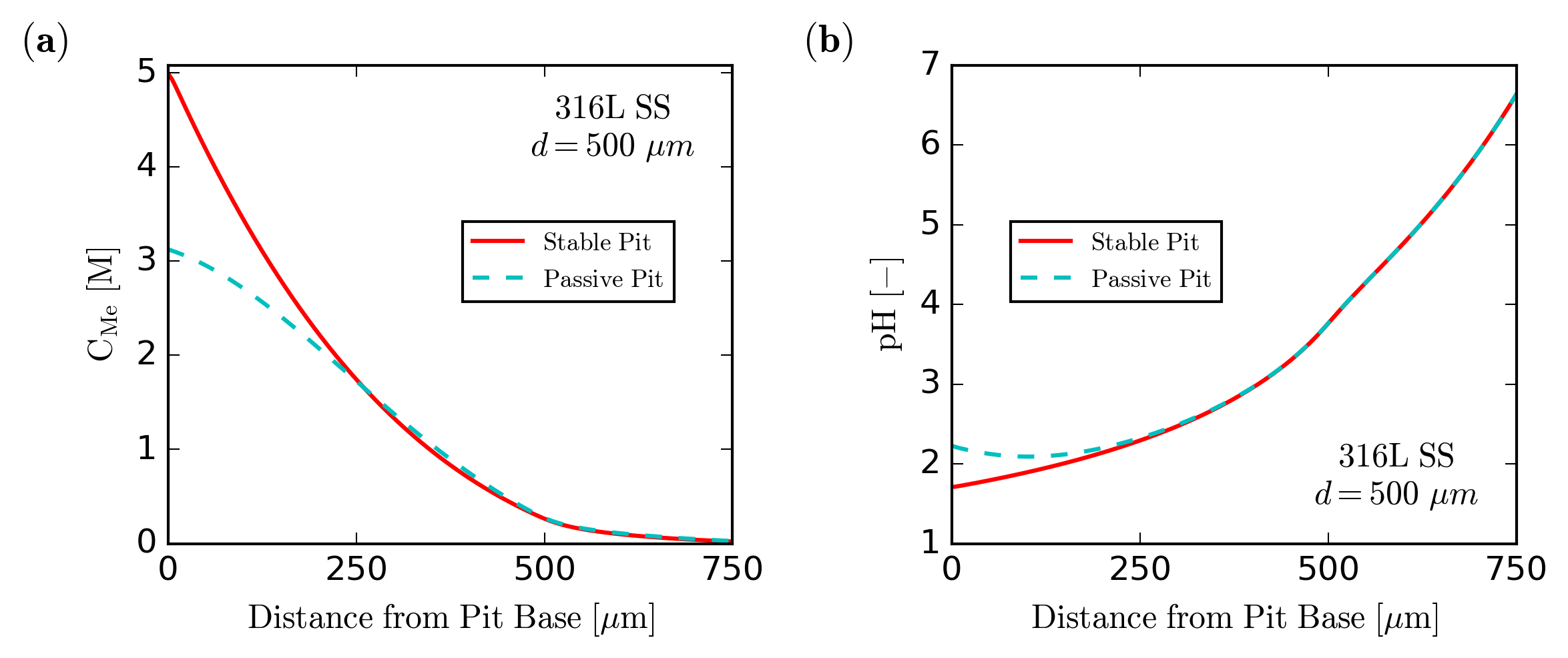}
  \caption{Comparison of the local chemistry inside the pit with depth $d=500 \mu$m at active (saturated state of Stage I) and passive (repassivated state of Stage II) states of pitting corrosion of 316L stainless steel wire within $0.6$ M NaCl under room temperature $T=25^{\circ}\mathrm{C}$: (a) Concentration of metal cation $C_\text{Me}$; (b) the pH.} \label{fig:stage2-candph}
\end{figure}

We first consider the pit with constant depth $d=500~\mu$m as it is about to repassivate. In \fref{fig:stage2-candph}, we compare the local chemistry inside the pit predicted by our model at the saturation and repassivation state. The dilution of metal cation at the corroding interface is in evidence and the concentration of the metal cation at the pit base decreases from $C_\text{sat}=5.02$ M to $61.16\%C_\text{sat}=3.07$ M, as shown in \fref{fig:stage2-candph}(a). There are two major mechanisms leading to the dilution of the metal cation at the pit base: Firstly, during the rapid polarization scan, the decrease of the applied electrical potential diminishes the current passing through the corroding surface. Therefore, the flux of metal cations entering the electrolyte via anodic dissolution at the pit base has become almost negligible at the onset of repassivation; Secondly, we know that ionic diffusion and electro-migration are controlled by the electrochemical potential gradient of the corresponding species \citep{nernst2009kinetik}. As long as the gradients of the concentration and/or the electrical potential exist, the species of metal cations and salts will continuously move out of the pit leading to the metal cation dilution during Stage II. 

From \fref{fig:stage2-candph}(b) we find that the pH near the corroding surface increases when the transition from pit stability to repassivation happens. Note that as mentioned earlier, a sufficiently low pH at the pit base is indicative that the aggressive chemistry necessary for continued pit stability is maintained. Therefore, an increase in pH at the pit base is naturally expected to take place for the pit to repassivate. Based on the qualitative analysis of the multi-species reactive transport model, we here propose two underlying mechanisms that can cause the increase of the pH near the pit base: Firstly, the dilution of metal cations breaks the local chemical equilibrium of reactions listed in Section \ref{sec:chemicalreaction}. To restore the chemical equilibrium, the hydrolysis of metal ions can occur in its backward direction, which consumes \ce{H^+} and produce metal cations. The consumption of hydrogen ion can then lead to the production of hydroxide ion through the water dissociation reaction and eventually cause the increase of the pH. We refer to this as the dilution-induced mechanism; Secondly, the suppressed local cathodic reaction (HER) will be re-activated once the applied electrical potential approaching to the repassivation potential. Therefore, the rate of the local cathodic reaction increases after the transition from pit stability to repassivation happens. As the major production of the local cathodic reaction, the hydroxide ion is supposed to dissolve into the pit solution from the corroding surface and then results in the pH rise near the pit base. We refer to this as the HER-induced mechanism.

\begin{figure}
  \centering
  \includegraphics[width=\textwidth,height=\textheight,keepaspectratio]{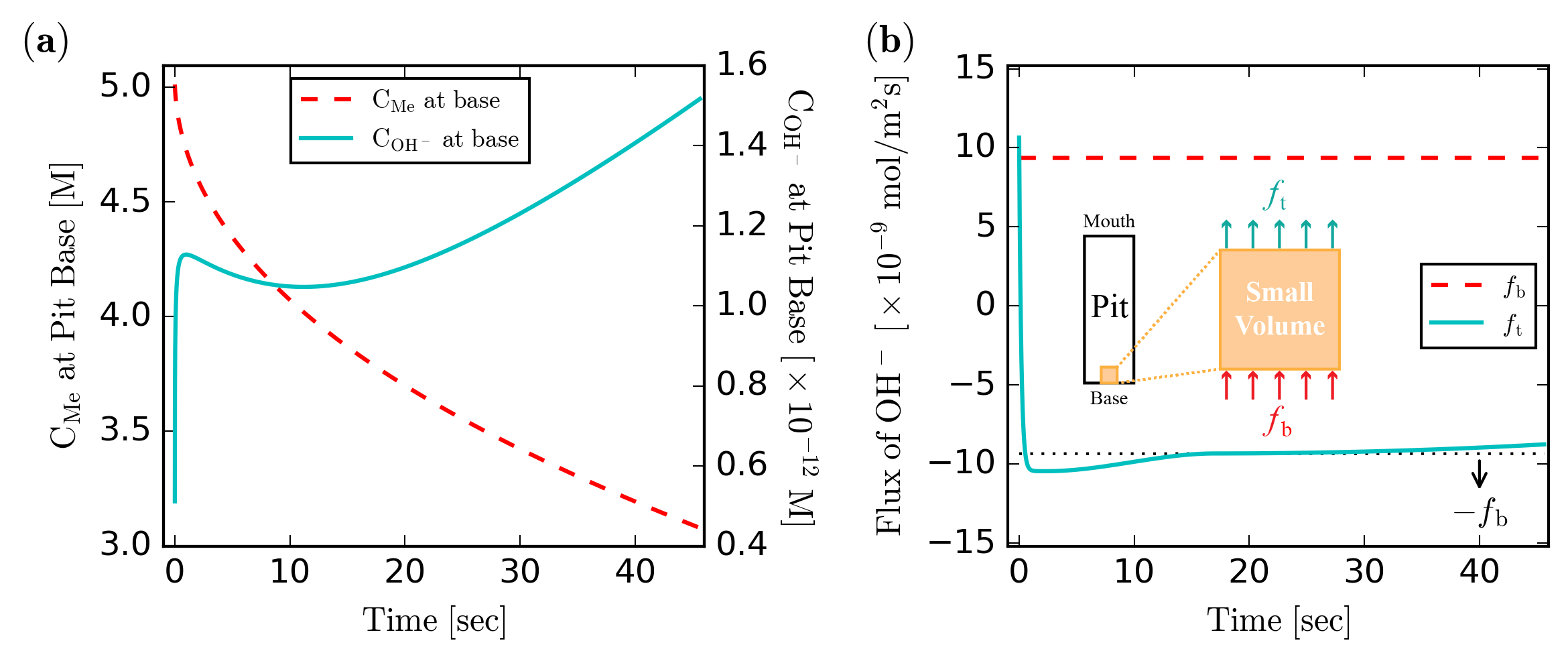}
  \caption{(a) Time dependence of metal cation and hydroxide ion concentration at the base of the pit with depth $d=500 \mu$m during Stage II under room temperature $T=25^{\circ}\mathrm{C}$ in $0.6$ M NaCl solution; (b) Schematic diagram of the pit base control volume and time dependence of the flux of hydroxide ion at the top and bottom surfaces of the volume.} \label{fig:stage2-evolution}
\end{figure}

To further determine the dominant mechanism that leads to the pH rise near the corroding surface, we here investigate the evolution of the critical chemistry at the pit base throughout the second stage, as shown in \fref{fig:stage2-evolution}. The metal cation concentration at the pit base is continuously being diluted during Stage II as evident from \fref{fig:stage2-evolution}(a). However, the concentration of hydroxide ion at the pit base changes non-monotonically (i.e., [\ce{OH^-}] increases initially followed by a decrease, but then again increases). According to the previous analysis about the dilution-induced mechanism, the necessary condition for it to dominate the pH rise is that the concentration variation of both metal cation and hydroxide ion with respect to time should be synchronous. Therefore, in this study, the dilution of metal cation cannot be the dominant mechanism resulting in the increase of the pH near the pit base. We next validate the HER-induced mechanism by calculating the flux of \ce{OH^-} near the corroding surface. As depicted in \fref{fig:stage2-evolution}(b), we take an extremely small control volume at the pit base with the dimension of $1\times1\ \mu\text{m}^2$. To estimate the amount of \ce{OH^-} inside the control volume, we compute the fluxes of \ce{OH^-} passing through the top ($f_\text{t}$) and bottom ($f_\text{b}$) surfaces as

\begin{equation}\label{eq:volumeflux}
	f_\text{t/b}=-D_{\text{OH}^\text{-}}\frac{\partial C_{\text{OH}^\text{-}}}{\partial y}+\frac{FD_{\text{OH}^\text{-}}C_{\text{OH}^\text{-}}}{R^*T}\frac{\partial \psi}{\partial y},
\end{equation}

\noindent where $y$ is the vertical coordinate originated from the pit base. Two right hand side terms of Equation \eref{eq:volumeflux} correspond to the fluxes of ionic diffusion and electro-migration, respectively. And species entering into the volume is counted positive flux, whereas species leaving from the volume is counted negative. The concentration of \ce{OH^-} at the pit base can then be approximated as

\begin{equation}\label{eq:cofohvstime}
	C_{\text{OH}^\text{-}}(t)=C_{\text{OH}^\text{-}}(0)+\int(f_\text{t}+f_\text{b})dt.
\end{equation}

\noindent As shown in \fref{fig:stage2-evolution}(b), the bottom flux of hydroxide ion $f_\text{b}$ is a positive constant, whereas the top flux of hydroxide ion $f_\text{t}$ decreases sharply from positive to negative at the beginning of the repassivation stage and then keep increasing. At Stage II, because the repassivation current density is assumed to be constant, we enforce a constant flux of \ce{OH^-} at the pit base as the boundary condition. Therefore, the bottom flux of hydroxide ion does not vary with time, which demonstrates that the change of \ce{OH^-} concentration at the pit base is actually controlled by the top flux. The time evolution of the top flux can be well understood by analyzing the direction of hydroxide ion diffusion and electro-migration. Initially, the concentration of \ce{OH^-} at the pit mouth is greater than that at the pit base, as shown in \fref{fig:stage2-candph}(b). Thus, the direction of hydroxide ion diffusion inside the pit is from the mouth to the base. Because the direction of the electrical current is always from the pit base to the pit mouth, the electrical potential is higher near the base and lower near the mouth. Considering the fact that \ce{OH^-} is a negatively charged ion, its electro-migration direction is the same as the diffusion direction. Therefore, \ce{OH^-} enters from both the top and bottom surface into the control volume at the beginning of Stage II, leading to the sudden increase of the hydroxide ion concentration at the pit base. However, such increase causes the concentration of \ce{OH^-} at the bottom surface to become greater than that at the top surface of the control volume. The diffusion direction of hydroxide ion inside the volume is then reversed (i.e., from the bottom to the top), which is contrary to the electro-migration direction. Consequently, the electro-migration flux is offset by the diffusion flux at the top surface. Hydroxide ion then turns to leave from the volume through the top surface because the diffusion gradually dominates the flux of \ce{OH^-}. No matter how the top flux increases afterwards, the bottom flux is the major source that results in the increase of the hydroxide ion concentration inside the pit base control volume. Because the bottom flux of \ce{OH^-} is generated from the local cathodic reaction, the HER-induced mechanism herein takes the dominant place that results in the pH rise at the pit base.

\begin{figure}
  \centering
  \includegraphics[scale=0.9,keepaspectratio]{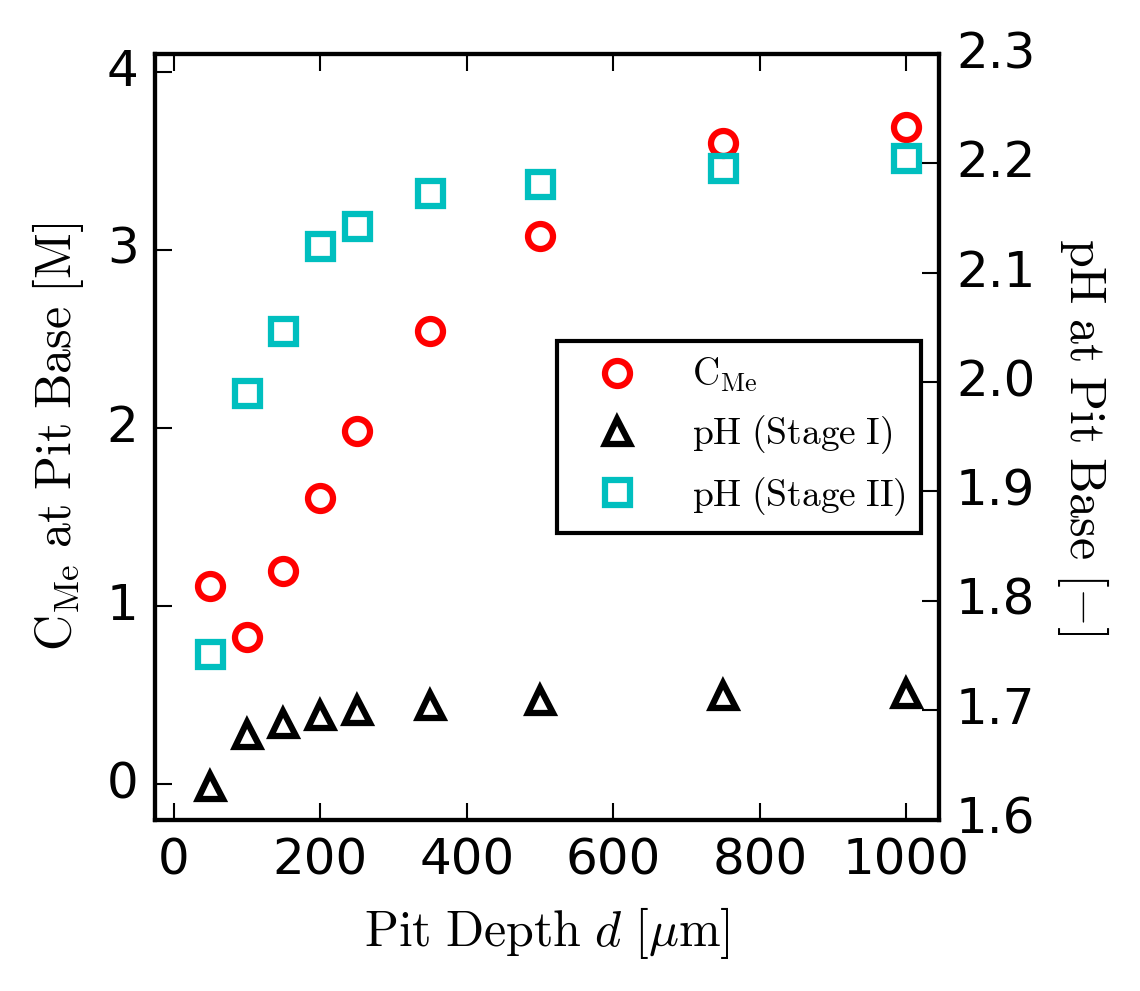}
  \caption{Numerical results of local chemistry inside the pit at the saturation (Stage I, replotted from \fref{fig:stage1-candph}(a) for comparison) and the repassivation (Stage II) states for pitting corrosion of 316L stainless steel wire within $0.6$ M \ce{NaCl} solution under room temperature $T=25^{\circ}\mathrm{C}$: the concentration of metal cation and the pH at the pit base with respect to different pit depths.} \label{fig:stage2-cphdepth}
\end{figure}

We next evaluate the local chemistry at the base of pits with different depths when repassivation happens. As demonstrated in \fref{fig:stage2-cphdepth}, because the metal cation is saturated initially, the dilution of metal cation at the pit base occurs at all depths when repassivation happens but to different levels. Given the short diffusion length of the shallow pits, the concentration gradient of metal cation is greater than that of the deep pits. The species transport inside shallow pits is faster and hence the concentrations of the metal cations at the base of shallow pits are lower than those at the base of deep pits. For sufficiently deep pits ($d/\diameter>8$), the critical surface concentration of metal cation is herein predicted to be in the range from $60\%$ to $75\%$ of saturation ($3.012$ to $3.765$ M).  Gaudet et al. estimated the critical solution chemistry of pits approaching repassivation to be about $60\%$ to $80\%$ of saturation based on multiple intermediate steady states that follow the salt film-free dissolution of 304 stainless steel artificial pits \citep{gaudet1986mass}. Ernst and Newman determined the lower bound value for this concentration to be around $60\%$ to $65\%$ of saturation from artificial pit experiments on 316L stainless steel \citep{ernst2007explanation}. Our prediction of the critical surface concentration is therefore consistent with these existing studies. Furthermore, we find that the pH of repassivation at the pit base is distinctly higher than the pH in the saturated state, as shown in \fref{fig:stage2-cphdepth}. As has been argued previously, the pH rise is mainly due to the production of hydroxide ion from the local cathodic reaction. Considering the fact that the flux of \ce{OH^-} is a constant regardless of the pit depth, the increase of the pH at the pit base is actually proportional to the elapsed time of the second stage $t_\text{II}$. However, our predicted critical pH of repassivation is about $2.2$, which is slightly less than the estimated critical pH of $2.65$ that enables oxide nucleation \citep{srinivasan2016evaluating}. Note that this simulated oxide nucleation pH is based on thermodynamics, which means any improved kinetic formulations can lead to the change of this estimation. Additionally, an underestimate of the ratio of the local cathodic current density to the anodic current density could have also resulted in a lower predicted repassivation pH as the \ce{OH^-} is mainly produced via the local cathodic reaction.

\section{Conclusions} \label{sec:conclusion}
In this study, we examine several critical electrochemical factors for artificial pit experiments of 316L stainless steel wire using a multi-species reactive transport model. The effects of viscosity and temperature on species diffusivity are considered and mathematically described using the Stokes-Einstein equation. The pit stability products and critical electrical potential ($E_\text{T}$ and $E_\text{rp}$) collected from experiments are applied to estimate the electrode current density and the elapsed time of the simulation. We consider two stages of simulation according to the change of the electrical current density during the artificial pit experiments: stable pitting under a salt film and film-free dissolution that transitions to repassivation. At the saturated-state of Stage I, the pH near the corroding surface is low enough to sustain the aggressive chemistry necessary for pit stability. The predicted specific resistance of the pit solution shows a good agreement with the experiment results. The effects of the temperature and the bulk concentration on pit electrochemical factors can be captured well using our model and are qualitatively consistent with existing experimental and numerical studies. At Stage II, the dilution of metal cation and the pH increase near the corroding surface are evident from the numerical results. The dilution of metal cation at the pit base is due to the transport of corresponding species and the decrease of the anodic current density. The critical surface concentration of metal cation is predicted to be in the range from $60\%$ to $75\%$ of saturation when repassivation happens. The pH rise at the pit base responsible for the onset of repassivation is a consequence of contribution by the local cathodic reaction, which is suppressed during the high-rate anodic dissolution in Stage I. In general, because we consider a more comprehensive mathematical model in this study, we are able to better understand the evolution of pit electrochemistry during two critical stages of artificial pit experiments.

\section*{Data availability}
\noindent The raw/processed data required to reproduce these findings cannot be shared at this time as the data also forms part of an ongoing study.

\section*{CRediT authorship contribution statement}
\noindent \textbf{Xiangming Sun:} Methodology, Software, Validation, Formal analysis, Investigation, Data Curation, Visualization, Writing - Original Draft. \textbf{Jayendran Srinivasan:} Methodology, Formal analysis, Investigation, Data Curation, Writing - Review \& Editing. \textbf{Robert G. Kelly:} Conceptualization, Resources, Writing - Review \& Editing. \textbf{Ravindra Duddu:} Conceptualization, Supervision, Project administration, Funding acquisition, Writing - Review \& Editing.

\section*{Declaration of Competing Interest}
\noindent The authors declare that they have no known competing financial interests or personal relationships that could have appeared to influence the work reported in this paper.

\section*{Acknowledgement}
\noindent This work was financially supported through start up funding from Vanderbilt University.

\newpage
\begin{appendices}
\section{Pit Stability Product}\label{appendix:pitstability}
\noindent Pit stability products obtained from artificial pit experiments of type 300 stainless steel wire in the \ce{NaCl} environment under different temperature as reported in $^\text{a}$ Srinivasan and Kelly \citep{srinivasan2016one}; $^\text{b}$ Katona et al. \citep{katona2019prediction}.

\begin{table}[h]
\centering
\begin{tabular}{ccccc}
\toprule
$C_\text{bulk}$ [M] & $25^{\circ}\mathrm{C}$ & $35^{\circ}\mathrm{C}$ & $45^{\circ}\mathrm{C}$ & $55^{\circ}\mathrm{C}$ \\ \midrule
$0.6$             & $0.90$$^\text{a}$                     & $1.06$$^\text{b}$                     & $1.10$$^\text{b}$                     & $1.45$$^\text{b}$                     \\
$1.0$             & $0.75$$^\text{b}$                     & $0.88$$^\text{b}$                     & $1.00$$^\text{b}$                     & $1.08$$^\text{b}$                     \\
$3.0$             & $0.54$$^\text{b}$                     & $0.60$$^\text{b}$                     & $0.70$$^\text{b}$                     & $0.90$$^\text{b}$                     \\
$5.5$             & $0.20$$^\text{b}$                     & $0.36$$^\text{b}$                     & $0.40$$^\text{b}$                     & $0.64$$^\text{b}$                     \\ \bottomrule
\end{tabular}%
\end{table}

\newpage
\section{Diffusion Coefficient}\label{appendix:diffusivity}
\noindent Reference values of species diffusion coefficients $D_\text{ref}$ under room temperature $298.15$ K.

\begin{table}[h]
\centering
\begin{tabular}{@{}cccccc@{}}
\toprule
Species & $D_\text{ref}$ [$10^{-9}\ \text{m}^2$/s] & Reference & Species & $D_\text{ref}$ [$10^{-9}\ \text{m}^2$/s] & Reference \\ \midrule
\ce{H^+} & $9.31$ & \citep{chang1998modeling} & \ce{OH^-} & $5.27$ & \citep{chang1998modeling} \\
\ce{Na^+} & $1.33$ & \citep{chang1998modeling} & \ce{Cl^-} & $2.03$ & \citep{chang1998modeling} \\
\ce{Fe^{2+}} & $0.824$ & \citep{gaudet1986mass} & \ce{Cr^{3+}} & $0.824$ & \citep{gaudet1986mass} \\
\ce{Ni^{2+}} & $0.824$ & \citep{gaudet1986mass} & \ce{Mo^{3+}} & $0.824$ & \citep{gaudet1986mass} \\
\ce{FeCl^{+}} & $0.824$ & \citep{gaudet1986mass} & \ce{FeCl_2} & $0.008$ & $-$ \\
\ce{FeOH^{+}} & $0.824$ & \citep{gaudet1986mass} & \ce{FeOH_2} & $0.008$ & $-$ \\
\ce{CrCl^{2+}} & $0.824$ & \citep{gaudet1986mass} & \ce{CrCl_2^+} & $0.824$ & \citep{gaudet1986mass} \\
\ce{CrCl_3} & $0.008$ & $-$ & \ce{CrOH^{2+}} & $0.824$ & \citep{gaudet1986mass} \\
\ce{CrOH_2^+} & $0.824$ & \citep{gaudet1986mass} & \ce{CrOH_3} & $0.008$ & $-$ \\
\ce{NiCl^{+}} & $0.824$ & \citep{gaudet1986mass} & \ce{NiCl_2} & $0.008$ & $-$ \\
\ce{NiOH^{+}} & $0.824$ & \citep{gaudet1986mass} & \ce{NiOH_2} & $0.008$ & $-$ \\
\bottomrule
\end{tabular}
\end{table}

\newpage
\section{Temperature-Pressure Dependence of Solution Viscosity}\label{appendix:viscosity}
\noindent Parameters $\zeta_{mn}$ calibrated from experiments \citep{aleksandrov2012viscosity} are used for the calculation of coefficient $a_i$ describing temperature-pressure dependence of solution viscosity as introduced in Equation \eref{eq:viscosityco}.

\begin{table}[h]
\centering
\begin{tabular}{c|c|cc}
\hline
$\zeta_{mn}$           & $n$                      & $m=0$                & $m=1$                \\ \hline
\multirow{5}{*}{$a_0$} & $0$  & $7.65306227\times10^{-1}$   & $-$                  \\
                       & $1$  & $-1.78087723$               & $ -1.39287121\times10^{-2}$  \\
                       & $2$  & $1.25374926$                & $1.50838568\times10^{-2}$    \\
                       & $3$  & $-1.61102861\times10^{-1}$  & $3.08379974\times10^{-3}$    \\
                       & $4$  & $-7.27978696\times10^{-2}$  & $ -5.27719410\times10^{-3}$  \\ \hline
\multirow{3}{*}{$a_1$} & $1$  & $9.84225044\times10^{-2}$   & $2.22367778\times10^{-2}$   \\
                       & $2$  & $2.43031313\times10^{-1}$   & $-3.68121974\times10^{-2}$  \\
                       & $3$  & $-2.05007685\times10^{-1}$  & $1.57724556\times10^{-2}$   \\ \hline
\multirow{3}{*}{$a_2$} & $1$  & $1.62231138\times10^{-1}$   & $-5.78612528\times10^{-3}$  \\
                       & $2$  & $-2.91448434\times10^{-1}$  & $9.45954845\times10^{-3}$   \\
                       & $3$  & $1.31605764\times10^{-1}$   & $-3.96672431\times10^{-3}$  \\ \hline
\multirow{3}{*}{$a_3$} & $1$ & $-2.08142179\times10^{-2}$   & $5.49316850\times10^{-4}$   \\
                       & $2$ & $3.80923122\times10^{-2}$    & $-9.01751419\times10^{-4}$  \\
                       & $3$ & $-1.66477653\times10^{-2}$   &  $3.77212002\times10^{-4}$  \\ \hline
\end{tabular}%
\end{table}

\newpage
\section{Calculation of Chemical Reaction Equilibrium Constants}\label{appendix:reaction}
\setcounter{equation}{0}
\renewcommand{\theequation}{\thesection.\arabic{equation}}
\noindent To better capture the temperature dependence of chemical reaction properties, we calculate the chemical reaction equilibrium constants $K_r$ following the calibrated formulation obtained from the Thermoddem database \citep{blanc2012thermoddem}
\begin{equation} \label{eq:chemicalconstant}
    \log_{10}K_r=b_0+b_1T+b_2T^{-1}+b_3\log_{10}(T)+b_4T^{-2},
\end{equation}

\noindent where $b_i$ are calibrated coefficients directly available from the database. We list $b_i$ for the equilibrium constants calculation of all chemical reactions considered in this work as below.

\begin{table}[h]
\centering
\begin{tabular}{cccccc}
\toprule
Chemical Reaction & $b_0$ & $b_1$ & $b_2$ & $b_3$ & $b_4$ \\ \midrule
\ce{Fe^{2+} + Cl^-\rightleftharpoons FeCl^+} & $808.65$ & $0.1318$ & $-45088.15$ & $-294.24$ & $2772595.80$ \\
\ce{Fe^{+} + 2Cl^-\rightleftharpoons FeCl_2} & $1614.41$ & $0.2611$ & $-93454.00$ & $-584.79$ & $5352140.98$ \\
\ce{Fe^{2+} + H_2O\rightleftharpoons FeOH^+ + H^+} & $200.44$ & $0.0301$ & $-13947.78$ & $-72.51$ & $648026.46$ \\
\ce{Fe^{2+} + 2H_2O\rightleftharpoons Fe(OH)_2 + 2H^+} & $333.01$ & $0.0508$ & $-21902.65$ & $-120.42$ & $927522.20$ \\
\ce{Cr^{3+} + Cl^-\rightleftharpoons CrCl^{2+}} & $1135.49$ & $0.1860$ & $-62020.19$ & $-413.21$ & $3569091.70$ \\
\ce{Cr^{3+} + 2Cl^-\rightleftharpoons CrCl_2^{+}} & $1774.63$ & $0.2887$ & $-97135.39$ & $-646.34$ & $5661676.10$ \\
\ce{Cr^{3+} + H_2O\rightleftharpoons CrOH^{2+} + H^+} & $233.14$ & $0.0362$ & $-15165.61$ & $-83.27$ & $837406.70$ \\
\ce{Cr^{3+} + 3H_2O\rightleftharpoons Cr(OH)_3 + 3H^+} & $547.11$ & $0.0791$ & $-33332.58$ & $-195.38$ & $1349304.75$ \\
\ce{Ni^{2+} + Cl^-\rightleftharpoons NiCl^+} & $796.76$ & $0.1293$ & $-44201.34$ & $-289.74$ & $2667655.45$ \\
\ce{Ni^{+} + 2Cl^-\rightleftharpoons NiCl_2} & $1567.32$ & $0.2550$ & $-90038.91$ & $-568.86$ & $5124711.71$ \\
\ce{Ni^{2+} + H_2O\rightleftharpoons NiOH^+ + H^+} & $236.97$ & $0.0362$ & $-16157.56$ & $-86.36$ & $945152.34$ \\
\ce{Ni^{2+} + 2H_2O\rightleftharpoons Ni(OH)_2 + 2H^+} & $329.16$ & $0.0511$ & $-21713.41$ & $-118.92$ & $979036.43$ \\
\ce{H^{+} + OH^{-} \rightleftharpoons H_{2}O} & $701.95$ & $0.1127$ & $-36168.25$ & $-253.60$ & $2423273.05$ \\
\bottomrule
\end{tabular}%
\end{table}

\end{appendices}

\clearpage
%\section*{\refname}
\bibliographystyle{elsarticle-num}
\renewcommand{\bibname}{References}
\bibliography{Ref}

\end{document}